\documentclass[aps,pre,twocolumn,superscriptaddress,nofootinbib]{revtex4-2}

\usepackage{amsmath}
\usepackage{mathrsfs}
\usepackage{amssymb}
\usepackage{bm} 
\usepackage{graphicx,grffile,textcomp}
\usepackage{dcolumn} 
\usepackage[usenames,dvipsnames]{xcolor}
\usepackage{multirow}
\usepackage{epstopdf}
\usepackage{mathtools} 
\usepackage{tabularx}
\usepackage{microtype}
\usepackage{comment}
\usepackage[version=4]{mhchem} 
\usepackage{array}
\usepackage{float}
\usepackage{wrapfig} 
\usepackage{url}
\usepackage{hyperref}
\usepackage{cleveref}
\usepackage[utf8]{inputenc}
\usepackage[T1]{fontenc}
\usepackage{bbold}

\hypersetup{colorlinks=true, linkcolor=blue, urlcolor=blue, citecolor=blue}

\definecolor{myred}{RGB}{230, 7, 77}

\definecolor{blue(ryb)}{rgb}{0.01, 0.28, 1.0}
\definecolor{jade}{rgb}{0.0, 0.66, 0.42}
\definecolor{cadmiumred}{rgb}{0.89, 0.0, 0.13}
\definecolor{darkviolet}{rgb}{0.58, 0.0, 0.83}
\definecolor{debianred}{rgb}{0.84, 0.04, 0.33}

\begin{document}

\title{Electrostatic interactions between charge regulated spherical macroions} 

\author{Hu Ruixuan}
\thanks{Co-first author}
\affiliation{School of Physical Sciences, University of Chinese Academy of Sciences, Beijing, 100049, China}
\author{Arghya Majee}
\thanks{Co-first author}
\affiliation{Max Planck Institute for the Physics of Complex Systems, 01187 Dresden, Germany}
\author{Jure Dobnikar}
\email{jd489@cam.ac.uk}
\affiliation{School of Physical Sciences, University of Chinese Academy of Sciences, Beijing, 100049, China}
\affiliation{CAS Key Laboratory of Soft Matter Physics, Institute of Physics, Chinese Academy of Sciences, Beijing, 100190, China}
\affiliation{Wenzhou Institute of the University of Chinese Academy of Sciences, Wenzhou, Zhejiang, China}
\affiliation{Songshan Lake Materials Laboratory, Dongguan, Guangdong 523808, China}
\author{Rudolf Podgornik}
\email{podgornikrudolf@ucas.ac.cn}
\affiliation{School of Physical Sciences, University of Chinese Academy of Sciences, Beijing, 100049, China}
\affiliation{CAS Key Laboratory of Soft Matter Physics, Institute of Physics, Chinese Academy of Sciences, Beijing, 100190, China}
\affiliation{Kavli Institute for Theoretical Sciences, University of Chinese Academy of Sciences, Beijing, 100049, China}
\affiliation{Wenzhou Institute of the University of Chinese Academy of Sciences, Wenzhou, Zhejiang 325011, China}
\affiliation{Department of Physics, Faculty of Mathematics and Physics, University of Ljubljana, Jadranska 19, 1000 Ljubljana, Slovenia}

\date{\today}

\begin{abstract}
We study the interaction between two charge regulating spherical macroions with dielectric interior and dissociable surface groups immersed in a monovalent electrolyte solution. The charge dissociation is modelled via the Frumkin-Fowler-Guggenheim isotherm, which allows for multiple adsorption equilibrium states. The interactions are derived from the solutions of the mean-field Poisson-Boltzmann type theory with charge regulation boundary conditions. For a range of conditions we find symmetry breaking transitions from symmetric to asymmetric charge distribution exhibiting annealed charge patchiness, which  results in like-charge attraction even in a univalent electrolyte -- thus fundamentally modifying the nature of electrostatic interactions in charge-stabilized colloidal suspensions.
\end{abstract}

\maketitle

\let\thefootnote\relax\footnotetext{\\ \sl{Dedicated to the legacy of fundamental contributions of Fyl Pincus to macromolecular electrostatics that have enlightened our understanding of complex (bio)molecular systems and have inspired and continue to inspire the soft-matter community.}}

\section{Introduction}

{\sl Electrostatic interactions} are a fundamental component of molecular forces in the colloid and nanoscale domains \cite{RevModPhys.82.1887}, dominating in particular various phenomena in biological and biomolecular context, as exemplified by their role in the physics of DNA \cite{Pincus2000} as well as other macromolecules \cite{Muthu2023}, physics of polyelectrolytes \cite{Pincus1976, Henle_2004,DOBRYNIN2005,Wong2006} and polyelectrolyte brushes \cite{Pincus1991, Pincus2001, Pincus2008b}, protein physics \cite{Zhou2018}, membrane physics \cite{Pincus_1990, Cev18,DAN200341,gao2019membrane}, physics of nucleic acids \cite{Cherstvy2011} and physics of viruses \cite{Siber2007,ZANDI20201}, with many outstanding contributions of Fyl Pincus. The nanoscale electrostatics has been modeled on various levels \cite{Dijkstra2021}, being standardly based on the Deryagin-Landau-Verwey-Overbeek (DLVO) paradigm \cite{Ver48}, and in particular on the mean-field formulation of the Poisson-Boltzmann (PB) theory \cite{Mar21,Vasileva2023,Blossey2023}, which has well defined limits of applicability, some pertaining to the model and some to the methodology \cite{Pincus2008a,Naji2013}. A very common variety of the PB theory is its linearized version, the Debye-H\"uckel (DH) theory \cite{Ohshima2012}, that in many cases allows for analytic calculations of interactions between macromolecular ions
\cite{Filippov2006,Ohshima2012,Siryk2021,Siryk2022}. 

Within the mean-field PB descriptions of the electrostatic interaction between two charged surfaces \cite{And95,dobnikar2004poisson,dobnikar2003many,Kolesnikov_2022}, one usually assumes a  constant surface charge density or constant surface potential boundary conditions \cite{Bor01,Mar21}, unless the surface charges are inhomogeneous \cite{Pincus2007,Pincus2008}. Although this simplifies the problem, most common naturally occurring nanoparticle and macromolecular surfaces of 
interest, e.g., hard colloidal particles \cite{Jianzhong2020b}, soft biological molecules including proteins \cite{Jianzhong2020}, membranes, and lipid vesicles \cite{Khunpetch2022, Khunpetch2023},  rarely satisfy either of them \cite{Pop10, Lun13}. They respond to their environment, especially to the presence of each other, in a way that modifies both the charge density as well as the surface potential, adjusting them according to the separation between them and the bathing solution conditions \cite{Tre16,Tre17}. 
This conceptual framework  with a long history is formally referred to as the \textit{charge regulation} \cite{Avn19}. 

Electrostatic interactions of two charge-regulated macroions in the case of the Langmuir adsorption isotherm \cite{kubincova2020interfacial} have been studied for two small point-like particles and their connection with the Kirkwood-Shumaker attractive  fluctuation interactions has been elucidated  \cite{Adzic2014, Adzic2015,Adzic2016}. In addition,  charge regulation framework has been applied to planar, chemically identical macroion surfaces with equal adsorption/desorption properties 
\cite{Beh99_JPCB, Bie04, Bor08, Sivan2022}, to chemically non-identical surfaces with different  adsorption/desorption properties
\cite{Cha76, Mcc95, Beh99_PRE, Cha06}, and to patchy surfaces with inhomogeneous charge distribution~\cite{boon2011charge}. Interactions in non-planar systems have been studied to a lesser extent  \cite{Krishnan2017,Behjatian2022} because the electrostatic potential inhomogeneities induced on curved surfaces need to be approached either by additional analytical approximations \cite{REINER199359,ETTELAIE1995131,Zypman2022} or by intensive numerical schemes. 

The important point of departure for us is that for chemically identical surfaces the surface charge densities have been without exception assumed to be equal on both surfaces based upon general symmetry considerations \cite{Sad99, Neu99, Tri99}. However, the underlying physical reasoning for such an assumption is not general and not based upon 
the detailed chemical nature of the surfaces bearing charge. 
The fact that two surfaces are chemically identical and, therefore, interact in the same 
way with the adjacent liquid, is \textit{not sufficient} to infer equal surface charge densities. 
In fact, the nature of the electrostatic fields follows from the   minimization of the relevant thermodynamic potential in the equilibrium state, yielding  
the surface charge densities without any additional symmetry assumptions \cite{Maj18}. Whether this minimum implies an equal or unequal surface charge densities of the interacting macroions may and, as will be shown below, does depend on the system under consideration.

Recently, the consequences of charge regulation in simple geometries, where the electrostatic potential depends only on a single coordinate and the corresponding electrostatic charges are homogeneous, were studied in detail~\cite{Maj18,Maj19,Maj20,Khunpetch2022}. The conclusion that emerged was that there exists a {\sl charge symmetry breaking} between the two interacting surfaces that can result in attractive interactions between nominally chemically identical surfaces. The sufficient condition for this symmetry breaking was identified as the existence of multiple (at least two) equilibrium states with different sign of the charges in the adsorption/desorption model, such as in the zwitterionic Frumkin-Fowler-Guggenheim model \cite{Bor01,Koopal2020}. The coupling between electrostatic interaction and the two surface dissociation minima can lead to a global minimum of the free energy, which shows a broken charge symmetry between the two interacting surfaces \cite{Maj18}.    

Here we study the electrostatic interaction between two identical {\sl spherical dielectric macroions} with zwitterionic surface dissociable groups, immersed in a monovalent electrolyte. 
The surface charge of the macroions is regulated by a dissociation process modeled by the Frumkin-Fowler-Guggenheim isotherm or, equivalently, the corresponding adsorption free energy.
We use the total free energy to derive the Euler-Lagrange equations, which are the Poisson-Boltzmann (PB) equation with charge-regulation (CR) boundary conditions. 
We obtain the solution with the numerical package COMSOL Multiphysics and evaluate the electrostatic potential, the total free energy and the corresponding force between the macroions.
We explore the nature of the macroion interactions, and particularly under what conditions the charge distribution on the two macroions can be asymmetric~\cite{Maj20} and/or spatially inhomogeneous~\cite{Behjatian2022}. 
The latter is important for spherical macroions with a dielectric core, where, unlike in the planar case, the electrostatic potential along the surface is not constant. 

We find that due to the interplay between
the ion adsorption and the spherical geometry of the macroions, solutions with inhomogeneous and asymmetric surface charge densities are possible,  even at separations that are much larger than the Debye length. 
The equilibrium surface charge densities on the two macroions can therefore differ in magnitude, sign and angular profile. 
The nature and extent of the inhomogeneities varies with the macroion separation, which can give rise to an overall attraction between the macroions. 
The mechanism for the onset of attraction is rationalized by an analytical treatment of a much simpler system of point particles immersed in an  electrolyte bathing solution.

 \section{Surface dissociation model}

 \begin{figure}[t!]
\includegraphics[width=\columnwidth]{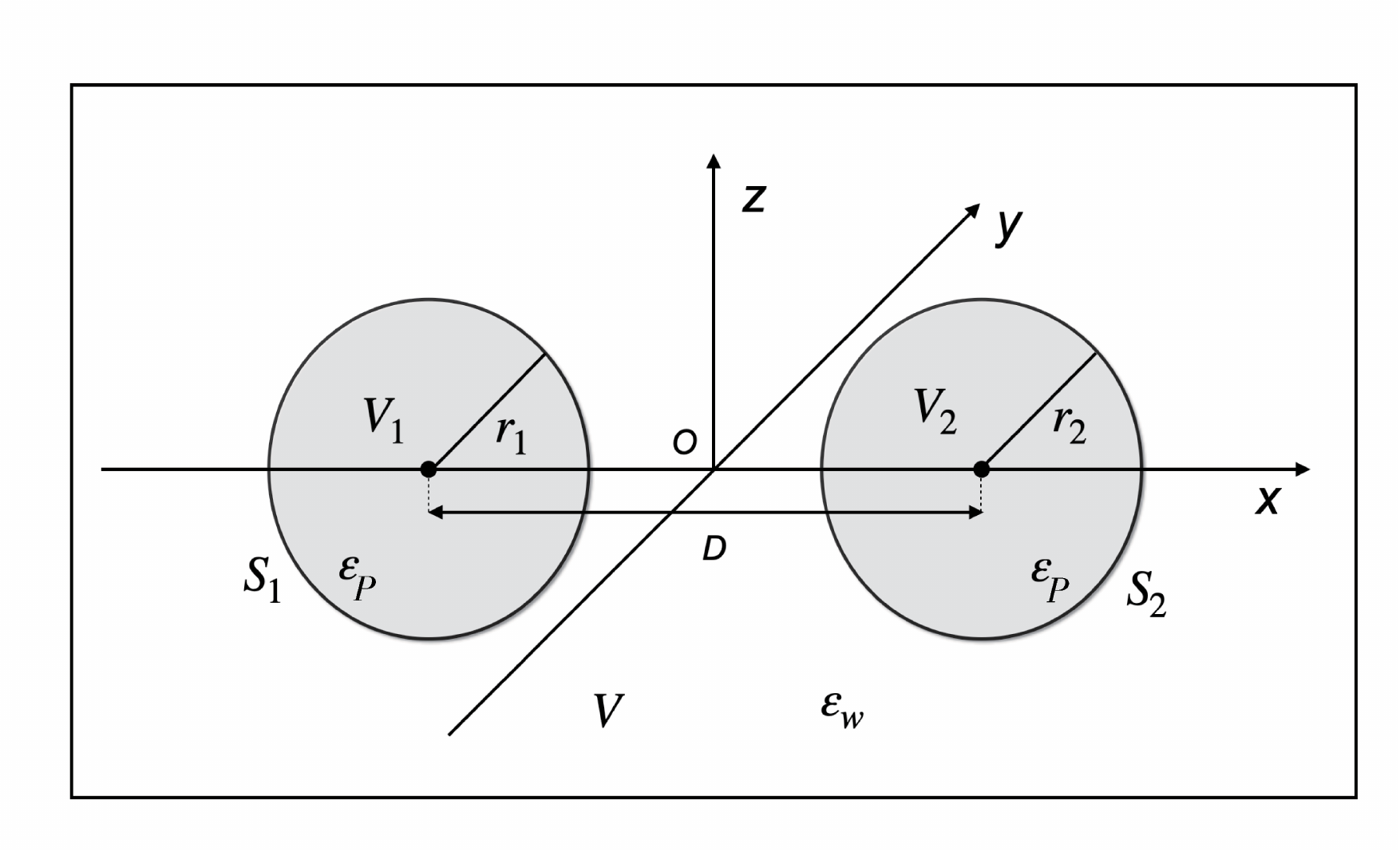}
\caption{Schematics of the problem of two spherical, charge regulated macroions of radii $r_1 = r_2 = R =50\,\mathrm{nm}$. The separation is $D$ and the dielectric constant values are $\varepsilon_W = 80$ and $\varepsilon_p = 4$ for the aqueous solution and the dielectric interior, respectively. In all calculations the ionic strength is taken as $10\,\mathrm{mM}$, corresponding to the Debye screening length of $3\,\mathrm{nm}$.} 
\label{Fig0}
\end{figure}

As shown in Fig.~\ref{Fig0}, we consider two spherical macroions of radius $R$ - whose surface charge depends on the charge-regulation model as introduced in \cite{Har06, Maj18} - that are suspended in a monovalent electrolyte solution. Each macroion surface contains a fixed number of dissociable zwitterionic surface groups, whose charge can be positive or negative. In addition, the charge on the spherical surface of the macroions is not necessarily homogeneous, but is a result of the dissociation equilibrium at each point along the surfaces. 

The surfaces are charge regulated through a dissociation process,  {characterized by an annealed surface degree of freedom within the model, $\phi$. By construction, $\phi\in[0,1]$.} If the area per site is $a^2$, with $n_0 = 1/a^2$, then the charge density $\sigma$ is identified as  
\begin{align}
    \sigma = e n_0\left(\phi-\frac{1}{2}\right)
    \label{sigma},
\end{align}
 {implying that as $\phi$ varies in the interval $0\leq\phi\leq 1$,} the charge density varies within a symmetric interval $-{\textstyle\frac12} {e}n_0\leq\sigma\leq{\textstyle\frac12} {e} n_0$,  
with $e>0$ being the elementary charge. This surface dissociation process is applicable to symmetric zwitterionic colloids with weak surface bound acids and bases that can be charged either positively or negatively \cite{Yuan2022},  {and in this case $\phi$ measures a dissociation imbalance between the acidic and basic moieties, in the sense that $\phi = 0$ pertains to negative sites being dissociated, while $\phi = 1$ to positive sites being dissociated. A more realistic model would need to take into account the fact that the charge interval in zwitterionic colloids need not be symmetric}.  

This zwitterionic colloid model is particularly relevant for lipid membranes composed of a mixture of cationic and anionic lipids \cite{Groves2004, Groves2009, Khunpetch2022} and even more so for polypeptides \cite{Maj20, Jav21}, where the dissociation can produce positively or negatively charged moieties, corresponding to cationic and anionic amino acid residues \cite{Finkel,Jianzhong2020}.

The dissociation isotherm for the charge regulation process can be obtained either from the chemical equilibrium mass action law equations or directly from the free energy corresponding to the dissociation, the path pioneered by Marcus \cite{Marcus1955} in the context of the reversible charging of macroions. The general connection between charge regulation and surface (dissociation) free energy has been elucidated in Ref.~\cite{Pod18}.   

Following Refs.~\cite{Har06, Maj18, Maj19, Maj20}, we base our macroion CR model on the Frumkin-Fowler-Guggenheim isotherm \cite{Koopal2020} of the macroion surface.  {This adsorption isotherm model pertains to a series of different phenomenological models invoked in the context of adsorption phenomena \cite{Abin-Bazaine22} and generalizes the Langmuir adsorption model by introducing the nearest neighbor interaction in addition to the mere adsorption energy.} The Frumkin-Fowler-Guggenheim model is defined with the phenomenological surface free energy density 
\begin{widetext}
\begin{eqnarray}
&&\int_{S} d^2{\bf r}~
f_{CR}( \phi) =  n_0k_BT \int_{S} d^2{\bf r}~
\Big( - \alpha\phi -\frac{1}{2} {\chi}\phi^2 
   + \phi\ln(\phi) 
   + (1-\phi)\ln(1-\phi)\Big).
   \label{modelfe}
\end{eqnarray}
\end{widetext}
The parameters $\alpha$ and $\chi$ are phenomenological and describe the non-electrostatic part of the ionic interactions at the macroion surface, the former quantifying the interaction of the ion with the surface, and the latter the interaction between ions already adsorbed to the surface. $\alpha \leq 0$ favours the adsorption of ions to the macroion surface, while $\chi \geq 0$ describes the correlation of adsorbed ions on the macroion surface inducing them to separate into domains.  The parameter $\chi$,  {which separates the Langmuir and the Frumkin-Fowler-Guggenheim models,} has in fact the same meaning as in the related regular lattice solutions theories (e.g., the Flory-Huggins theory \cite{Ter02}),  describing  the  short-range interactions between nearest  neighbor adsorption sites on the macroion surface \cite{Avn20}.  {In the case of $\chi = 0$ the Frumkin-Fowler-Guggenheim model is reduced directly to the Langmuir model.}

The  {surface dissociation products} can be either electrostatic potential determining electrolyte ions, or the pH determining protons as in the case of the protein and lipid (de)protonation dissociation.  In the case of (de)protonation reaction, the dependence of $\alpha$ on the bulk pH is model specific \cite{Avn18}, but on the mean-field level one can explicitly identify $\alpha=(\mathrm{pK}-\mathrm{pH})\ln 10$, 
where $\mathrm{pH} = -\log_{10}[\mathrm{H}^{+}]$, with $[\mathrm{H}^{+}]$ being the proton concentration in the bulk and $\mathrm{pK}$ is the dissociation constant of the deprotonation reaction \cite{Nin71}.  {For the (de)protonation dissociation reaction the dissociation isotherm in the Frumkin-Fowler-Guggenheim model would reduce to the Henderson-Hasselbalch equation if one takes $\chi =0$}.  In general, the proton concentration needs to be identified as the proton activity \cite{Holm1,Holm2, Levin2020,Levin2022}.

The model free energy Eq.~\eqref{modelfe} describes only the  non-electrostatic interactions at the macroion surface and thus cannot describe the dissociation equilibrium. In order to incorporate the electrostatic interactions on the continuum level one needs to add the electrostatic surface free energy density, that has the simple form $\sigma \psi $, $\psi$ being the surface electrostatic potential, and consequently the total surface free energy density can be written as
\begin{widetext}
\begin{eqnarray}
&&
\sigma\psi + f_{CR}(\phi)  =  n_0 k_BT  
\Big( \beta e(\phi-{\textstyle\frac12})\psi -\alpha\phi-\frac{1}{2} \chi \phi^2 + \phi\ln\phi+(1-\phi)\ln(1-\phi)\Big),  
\label{modelequ2}
\end{eqnarray}
\end{widetext}
with $\sigma$ defined in Eq.~\eqref{sigma}. The surface dissociation corresponding to the above model describes also the charge regulation (CR) after being coupled to the free energy of the mobile ions in the bathing solution, as we do next.

The Frumkin-Fowler-Guggenheim isotherm model of macroion surface dissociation Eq.~\eqref{modelequ2} was applied to lamellar-lamellar phase transition in a charged surfactant system \cite{Har06} and 
a good correspondence with experiments was obtained for the didodecyldimethylammonium 
chloride (DDACl) data with $\alpha = -3.4$, $\chi=14.75$, and for the didodecyldimethylammonium 
bromide (DDABr) data with $\alpha =-7.4$ and $\chi=14.75$, see Ref. \cite{Har06} for details. The same model was successfully applied also to other systems, see e.g., \cite{Fink2017,Fink2019}. 

In what follows we will use both $\alpha$ as  well as $\chi$ as purely phenomenological interaction  parameters, quantifying the adsorption energy in  the surface (de)protonation reactions and the 
nearest-neighbor surface energy of filled surface adsorption sites.

\section{Total free energy}

The total free energy of the system, $\cal E$, is assumed to be composed of three parts stemming from: (i) the electrostatic free energy of the mobile ions in the space outside the macroions, $V$, (ii) the electrostatic energy of the dielectric interior of the macroions, $V_1$ and $V_2$, and (iii) the surface free energies of charge dissociation on the surfaces of the two macroions, $S_1$ and $S_2$, thus
\begin{eqnarray}
 {\cal E} &=& {\cal F}_{ES} [V] + {\cal F}_{ES} [V_1] + {\cal F}_{ES} [V_2] + {\cal F}_{CR} [S_1] + {\cal F}_{CR} [S_2], \nonumber\\
 ~
\end{eqnarray}
with $V_1$ the volume of the first sphere with surface area $S_1$ and $V_2$ the volume of the second sphere with surface area $S_2$. $V$ is the volume in between the two spheres. ${\cal F}_{CR}$ is given by the model Eq.~\eqref{modelequ2}.

The Helmholtz free-energy of a monovalent electrolyte is given by the standard mean-field Poisson-Boltzmann expression \cite{Mar21}
\begin{widetext}
\begin{eqnarray}
\label{b13}
{\cal F}_{ES} [V] &=& \int_V d^3{\bf r} \left[ -\frac{1}{2} {\varepsilon_0\varepsilon_w}
\left( \nabla\psi({\bf r}) \right)^2 +  \sum_{i=1}^2  q_i n_i({\bf r}) \psi({\bf r})  + k_BT \sum_{i=1}^2  \Bigg( n_i({\bf r})\ln\left[ n_i({\bf r}) a^3 \right] - n_i({\bf r})  \Bigg) + \rho_0({\bf r}) \psi({\bf r}) \right],
\end{eqnarray}
\end{widetext}
where $n_{1,2}$ are the cation, $q_1 = e_0$, and anion, $q_2 = -e_0$, densities, $a$ is their size, with $\rho_0$ being the  external charge density.  {In the case considered, $\rho_0$ corresponds to the surface charge density resulting from the dissociation process as described by the zwitterionic dissociation model, Eq.~\eqref{sigma}.}

The sum of the first two terms in the above equation is equal to the electrostatic energy and the third one is the ideal gas entropy. In writing the above free energy we assumed that the concentration of $\mathrm{H}^{+}$ and $\mathrm{OH}^{-}$ ions, resulting from acid/base  dissociation and related to the bulk pH value, is (much) smaller than the concentration of electrolyte ions.

In addition, the electrostatic free energy inside the dielectric cores of both macroions 
\begin{eqnarray}
{\cal F}_{ES} [V_{1,2}] = \!\!\int_{V_{1,2}}\!\!\!\!\!\!\!d^3{\bf r} \Big[ -\frac{1}{2} {\varepsilon_0\varepsilon_P}
\left( \nabla\psi({\bf r}) \right)^2 + \rho_0({\bf r}) \psi({\bf r}) \Big]~~~
\end{eqnarray}
is given by the electrostatic energy term only, where $\varepsilon_{w,P}$ are the dielectric constants of water and the macroion dielectric interior, respectively.  The following should be noted in connection with the above expressions \cite{Schwinger1998}: the electrostatic field energy, proportional to the square of the field, comes with a minus because after the minimization the sum of this term and the external charge density term gives back correctly the positive definite free energy, see \cite{Maggs2016}.

The above electrostatic free 
energy  can be written in many equivalent forms \cite{Theodoor1990, Lamm2003, Bie04, Majee2016, Bebon2020} and we chose the one that seems the best suited for our calculation and expresses the free energy entirely as a functional of the local electrostatic potential \cite{Maggs2016}, containing two terms: the field energy proportional to $(\nabla \psi)^2$ and osmotic pressure of mobile ions, proportional to $\cosh{\beta e_0 \psi}$. The total free energy then assumes the form
\begin{widetext}
\begin{eqnarray}
{\cal E} [\psi({\bf r}), \phi({\bf r})] &=& - \int_{V} d^3{\bf r}\left( \frac{1}{2}\varepsilon_0  \varepsilon_w\left( \nabla \psi({\bf r})\right)^2 + \frac{k_BT~\kappa_D^2}{4 \pi \ell_B} \cosh{\beta e_0 \psi({\bf r}) }\right) - \nonumber\\
&& - \frac{1}{2} \varepsilon_0  \varepsilon_P \int_{V1} d{\bf r} \left( \nabla \psi({\bf r})\right)^2 + \int_{S_1} d^2{\bf r}~ f_{CR}(\psi, \phi)  - \frac{1}{2} \varepsilon_0  \varepsilon_P \int_{V2} d{\bf r} \left( \nabla \psi({\bf r})\right)^2 + \int_{S_2} d^2{\bf r}~ f_{CR}(\psi, \phi).~~~~~~~
\label{equ1}
\end{eqnarray}
\end{widetext}
Above we also introduced the inverse Debye screening length squared as
\begin{eqnarray}
    \kappa_D^2 \equiv 4\pi \ell_B n,
    \label{Debyescreen}
\end{eqnarray}
implying that the screening length is $\lambda_D = \kappa_D^{-1} = 0.304/\sqrt{n}$ in nm and $n$ in molar units. $\ell_B $ is the Bjerrum length, equal to the distance at which two unit charges interact with thermal energy $k_BT$ and given by  $\ell_B = e_0^2/4\pi\varepsilon_w\varepsilon_0 k_BT$ (in water at room temperature, the value is $\ell_B\approx 0.7\,\mathrm{nm}$) while $n$ is the bulk concentration of the electrolyte ions.  {The above form of the Debye screening length first of all implies a grand canonical equilibrium with a fixed chemical potential of ions equal to its value in the bulk salt reservoir, as well as a negligible concentration of the surface dissociation products, be it potential determining ions or (de)protonation products. In case this constraint is not met, a more detailed description of the various charged species would be needed \cite{Lan20}.}

Minimization of the free energy with respect to the electrostatic potential and with respect to the surface dissociation fraction then gives six Euler-Lagrange equations: the PB equation in $V$, the Poisson equation in $V_1$ and $V_2$, plus two Gaussian boundary conditions and two charge regulation equations on the surfaces of the two macroions $S_1$ and $S_2$. For the electrostatic potential minimization this yields the PB equation in $V$ 
\begin{eqnarray}
\nabla^2  \left(\beta e_0 \psi({\bf r})\right) = 
\kappa_D^2 \sinh{\beta e_0 \psi({\bf r}) } 
\label{pbequ}
\end{eqnarray}
and the Poisson equation in $V_1$ and $V_2$
\begin{eqnarray}
\nabla^2  \psi_{1,2}({\bf r}) = 0,
\label{pequ}
\end{eqnarray}
with the Gaussian boundary conditions at $S_1$ and $S_2$ amounting to 
\begin{eqnarray}
&& \frac{\partial f_{CR}(\psi, \phi)}{\partial \psi}\bigg\vert_{S_1} = \varepsilon_0 {\bf n}_1\cdot \left( \varepsilon_w\nabla \psi - \varepsilon_P\nabla\psi_1\right)  = \nonumber\\
&& ~~~~~~~~~~~~~~~~~~~ = \sigma_1 = n_0 e \left(\phi_1 - \frac{1}{2}\right) \nonumber\\ && \frac{\partial f_{CR}(\psi, \phi)}{\partial \psi}\bigg\vert_{S_2} = \varepsilon_0 {\bf n}_2\cdot \left( \varepsilon_w\nabla \psi - \varepsilon_P \nabla\psi_2\right)= \nonumber\\ 
&& ~~~~~~~~~~~~~~~~~~~ = \sigma_2 = n_0 e \left(\phi_2 - \frac{1}{2}\right),
\label{Gauss12}
\end{eqnarray}
where ${\bf n}_1$ points into volume 1 and ${\bf n}_2$ into volume 2, while the minimization with respect to the two dissociation fractions yields two charge regulation equations 
\begin{eqnarray}
\frac{\partial f_{CR}(\psi, \phi)}{\partial \phi}\bigg\vert_{S_1} = 0 \quad {\rm and} \quad \frac{\partial f_{CR}(\psi, \phi)}{\partial \phi}\bigg\vert_{S_2} = 0
\label{surfmin}
\end{eqnarray}
which taking into account the form of the CR surface free energy Eq.~\eqref{modelequ2} yields the following two Frumkin-Fowler-Guggenheim charge dissociation isotherms 
\begin{eqnarray}
&&\phi_1 = \left( 1 + e^{+\beta e \psi_1 - \chi_1 \phi_1 - \alpha_1}\right)^{-1} \nonumber\\
&&{\rm and} \qquad \phi_2 = \left( 1 + e^{+\beta e \psi_2 - \chi_2 \phi_2 - \alpha_2}\right)^{-1}.
\label{dissoc12}
\label{bcoqrwyu}
\end{eqnarray}
Clearly these two equations reduce to the surface Langmuir charge dissociation isotherms \cite{Nin71} if the interaction parameter $\chi$ vanishes for both surfaces. Finally combining the PB equation Eq.~\eqref{pbequ} and the Poisson equations Eq.~\eqref{pequ} with the Gaussian boundary condition Eqs.~\eqref{Gauss12} and the charge regulation dissociation equilibrium Eqs.~\eqref{dissoc12} yields a self-consistent system of equations that needs to be solved.

In addition one has to take into account that the electrostatic potential $\psi$ is of course continuous across the two boundaries $S_1$ and $S_2$ of the macroions. 

Alternatively the electrostatic part of the free energy can be equivalently written in the form of a {\sl Casimir charging process} that reduces the free energy to surface integrals \cite{verwey48a}:
\begin{widetext}
\begin{eqnarray}
{\cal F}_{ES} [\psi({\bf r})] &=& - \int_{V} d^3{\bf r}\left( \frac{1}{2}\varepsilon_0  \varepsilon_w\left( \nabla \psi({\bf r})\right)^2 + \frac{k_BT~\kappa_D^2}{4 \pi \ell_B} \cosh{\beta e_0 \psi({\bf r}) }\right)  + \int_{S_1}\!\!d^2{\bf r}~ \sigma_1\psi_1 + \int_{S_2}\!\!d^2{\bf r}~ \sigma_2\psi_2 = \nonumber\\
&=& \oint_{S_1} d^2{\bf r} \int_0^{\sigma_1} \psi(\sigma_1) d\sigma_1 + \oint_{S_2} d^2{\bf r} \int_0^{\sigma_2} \psi(\sigma_2) d\sigma_2 =   \oint_{S_1} d^2{\bf r} f_{ES}(\sigma_1) + \oint_{S_2} d^2{\bf r} f_{ES}(\sigma_2),
\end{eqnarray}
\end{widetext}
where the Casimir charging process at the surfaces, corresponding to the second line and described first in the Verwey-Overbeek classic \cite{Ver48, Theodoor1990},  yields a particularly simple - if implicit - form of the electrostatic free energy. Above,  $\psi(\sigma_1) = \psi_1$ and $\psi(\sigma_2) = \psi_2$ represent the solutions of the PB equation at the surfaces of the two macroions. We will see that this form of the electrostatic free energy allows for some further simplifications.

Using the Casimir charging form of the electrostatic free energy then yields the complete free energy in the form
\begin{eqnarray}
&&{\cal E} [\psi({\bf r}), \phi({\bf r})] = \oint_{S_1} \!\!\!d^2{\bf r} \int_0^{\sigma_1} \!\!\!\psi_1 d\sigma_1 + \int_{S_1} \!\!\!d^2{\bf r}f_{CR}( \phi)\bigg\vert_{S_1} + \nonumber\\
&& ~~~~~~~+ \oint_{S_2} \!\!\!d^2{\bf r} \int_0^{\sigma_2} \!\!\!\psi_2 d\sigma_2  + \int_{S_2}\!\!\! d^2{\bf r}~ f_{CR}( \phi)\bigg\vert_{S_2},~~~
\label{bgfhjdsk}
\end{eqnarray}
where we recall the definition in Eq.~\eqref{modelfe}. Note that $f_{CR}( \phi)$ and $f_{CR}( \psi, \phi)$ in Eq.~\eqref{modelequ2} differ in the electrostatic term included in the latter but not the former. The free energy in the Casimir form contains the complete free energy including  the PB terms for the electrooyte ions, and not only the surface CR  interactions.

Minimization of free energy Eq.~\eqref{bgfhjdsk} can then be cast in an alternative form
\begin{eqnarray}
\frac{\partial f_{ES}(\sigma_i)}{\partial \sigma_i} \frac{\partial \sigma_i}{\partial \phi_i} + \frac{\partial f_{CR}(\phi_i)}{\partial \phi_i} = 0 
\end{eqnarray}
wherefrom
\begin{eqnarray}
\beta e \psi_i  -  \chi_i\phi_i  + \ln{\frac{\phi_i}{1-\phi_i}} - \alpha_i =0 
\end{eqnarray}
for $i=1,2$, which is completely equivalent to the Frumkin-Fowler-Guggenheim isotherm in Eqs.~\eqref{bcoqrwyu} above, thus justifying the alternative formulation of the free energy. The above charge dissociation isotherm is only valid within the mean-field approximation limit. We note that in Eq.~\eqref{bgfhjdsk} the electrostatic part also contains terms dependent on $\phi$ and thus to the lowest order renormalizes the purely non-electrostatic parameters $\alpha, \chi$.

The free energy derived above is in a form that is not fully suitable for numerical calculations for the reason that it contains derivatives of the electrostatic potential which, in standard numerical software, are less accurate than the electrostatic potential values themselves.  {In Appendix \ref{Sec:A-App} we thus derive an alternative expression containing only the values of the electrostatic potential that we use in the actual  numerical calculations.}

In addition to the free energy we can also calculate the interactions in the system {\sl via} the stress tensor \cite{Budkov_2022,brandyshev2023}, that can be obtained in the standard form
\begin{eqnarray}
    \sigma_{ij} = \sigma_{ij}({\rm Maxwell}) + \delta_{ij} ~p({\rm van't~Hoff})
\end{eqnarray}
where the two components refer to the Maxwell electrostatic stress tensor and the van't Hoff expression for the osmotic pressure of the electrolyte ions. Explicitly this yields
\begin{widetext}
\begin{eqnarray}
    \sigma_{ij}({\bf r}) = \varepsilon_0 \Big( \nabla_i\psi({\bf r}) \nabla_j \psi({\bf r}) - \frac{1}{2} \delta_{ij} \left(\nabla\psi({\bf r})\right)^2\Big) - \delta_{ij} ~2 k_BT n \left(\cosh{\beta e_o \psi({\bf r})} - 1\right).
\end{eqnarray}
\end{widetext}
The force on a volume enclosed by a surface $S$ with a local normal ${\bf n} \equiv n_i$ is then given by
\begin{equation}
    {\cal F}_i = \oint_S \sigma_{ij} n_j~ dS. 
    \label{forcecal}
\end{equation}
The most convenient choice of the surface $S$ for two identical macroions is the midplane that encloses one of the macroions at infinity, so that the above surface integral would yield the force on that macroion. The most important feature of the force calculation via the stress tensor for an appropriate choice of the integration surface is that it does not contain any surface terms and its form is universal. 

This concludes our formal derivations and we now turn our attention to numerical results.   

\section{Results}

Numerical solution of the problem of two spherical CR macroions with free energy given by Eq.~\eqref{equ1},  {which is reduced to solving equations Eq.~\eqref{pbequ} to Eq.~\eqref{bcoqrwyu}, with the interaction forces evaluated with the help of Eq.~\eqref{forcecal}}, is computed with finite-element software package COMSOL Multiphysics. The geometry is shown on Fig.~\ref{Fig0}. We fix the macroion radii $r_1 = r_2 = R = 50\,\mathrm{nm}$ and vary their separation $D$. The dielectric constant values for the aqueous solution and the dielectric interior are $\varepsilon_W = 80$ and $\varepsilon_p = 4$, respectively. In all calculations the ionic strength is taken as $10\,\mathrm{mM}$, corresponding to the Debye screening length of $3\,\mathrm{nm}$. The parameter space of our problem is spanned by $\chi$ and $\alpha$ characterizing the surface adsorption/dissociation process, and by macroion separation. We do not intend to exhaustively explore the entire parameter space but choose to focus on the subset of the parameters and discuss the onset of asymmetry and inhomogeneity in the solutions.

\begin{figure*}[t!]
\includegraphics[width=14cm]{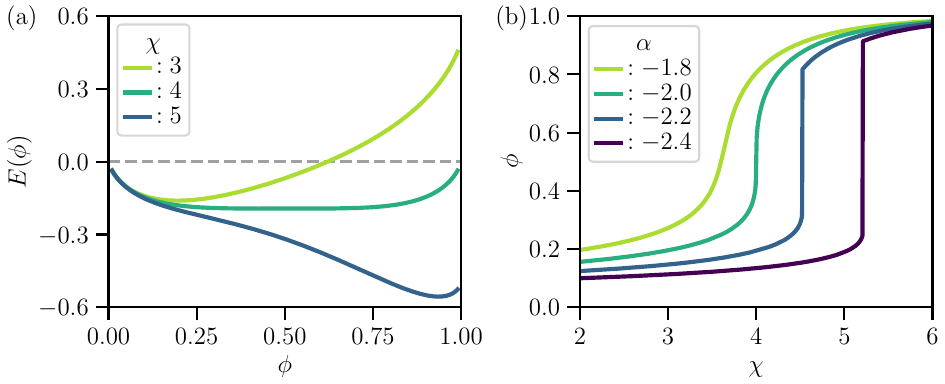}
\caption{(a) The CR free energy of the zwitterionic surface dissociation model Eq.~\eqref{modelfe} at $\alpha = -2$ for three different values of $\chi = 3,4,5$ (top to bottom curves), showing a discontinuous transition between two minima at $\phi \simeq 0$ and $\phi \simeq 1$. (b) The corresponding equilibrium $\phi$ as a function of $\chi$ for different choices of $\alpha = -1.8, -2, -2.2, -2.4$ (left to right curves). There obviously exists a "critical point" given by the combination $\alpha = -2.$ and $\chi = 4.$, therefore $\alpha = - {\textstyle\frac12} \chi$, delimiting the "subcritical region" $\alpha \geq - {\textstyle\frac12}\chi$ with a discontinuous dependence of $\phi$, and a continuous dependence in the "supercritical region" $\alpha < - {\textstyle\frac12}\chi$.} 
\label{Fig1}
\end{figure*}

\subsection{Single surface with CR}

We first briefly analyze the "equation of state" of the zwitterionic dissociation model, {\sl i.e.}, the dependence of the equilibrium value of $\phi$ on the parameters of the system for a single planar surface without considering the electrostatic effects, {\sl i.e.}, obtained by minimizing the CR Frumkin-Fowler-Guggenheim free energy, Eq.~\eqref{modelfe}. Clearly, for $\chi=0$, the problem reduces to the standard Langmuir adsorption, while introducing the adsorbed ion correlations ($\chi>0$) qualitatively changes the behaviour. As shown in Fig.~\ref{Fig1}, the CR free energy develops a second minimum for non-zero $\chi$,  {meaning that there are two locally stable states corresponding to two charged states {\sl via} Eq.~\eqref{sigma}, one with $\phi\approx 0$ corresponding to a predominantly negatively charged surface, and one with  $\phi\approx 1$, corresponding to a predominantly positively charged surface}. 
The relative stability of these two states depends on the values of $\alpha$ and $\chi$, with the "critical dissociation isotherm" corresponding to $\alpha = - {\textstyle\frac12}\chi$, with  a discontinuous transition between the two minima in the "subcritical region" $\alpha \geq - {\textstyle\frac12}\chi$ , and a continuous dependence in the "supercritical region" $\alpha < - {\textstyle\frac12}\chi$.

One could consider a slightly less complex version of the problem by adding an approximate description of the electrostatic effects in the minimization~\cite{Maj18}, however, we prefer to address the full PB CR problem.
The important feature of this analysis is that the Frumkin-Fowler-Guggenheim CR model exhibits multiple free energy minima and electrostatic interaction then modifies these minima and couples the interacting CR surfaces. This leads to a rich separation dependence of the equilibrium state that was noticed on a more phenomenological level not involving charge regulation  before \cite{Podgornik1989, Kornyshev1992}. We analyze the details of this coupling next.

\begin{figure*}[t!]
\includegraphics[width=14cm]{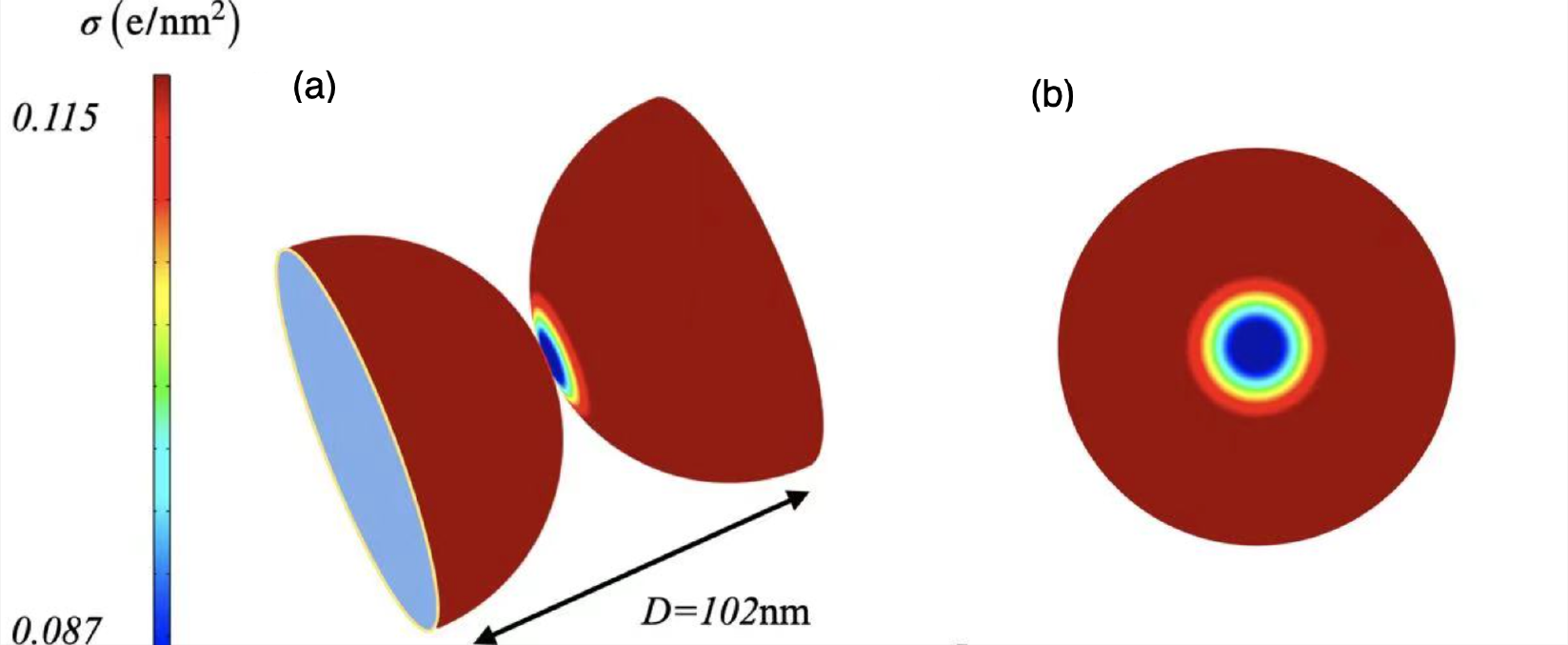}\\
~\\
\includegraphics[width=14cm]{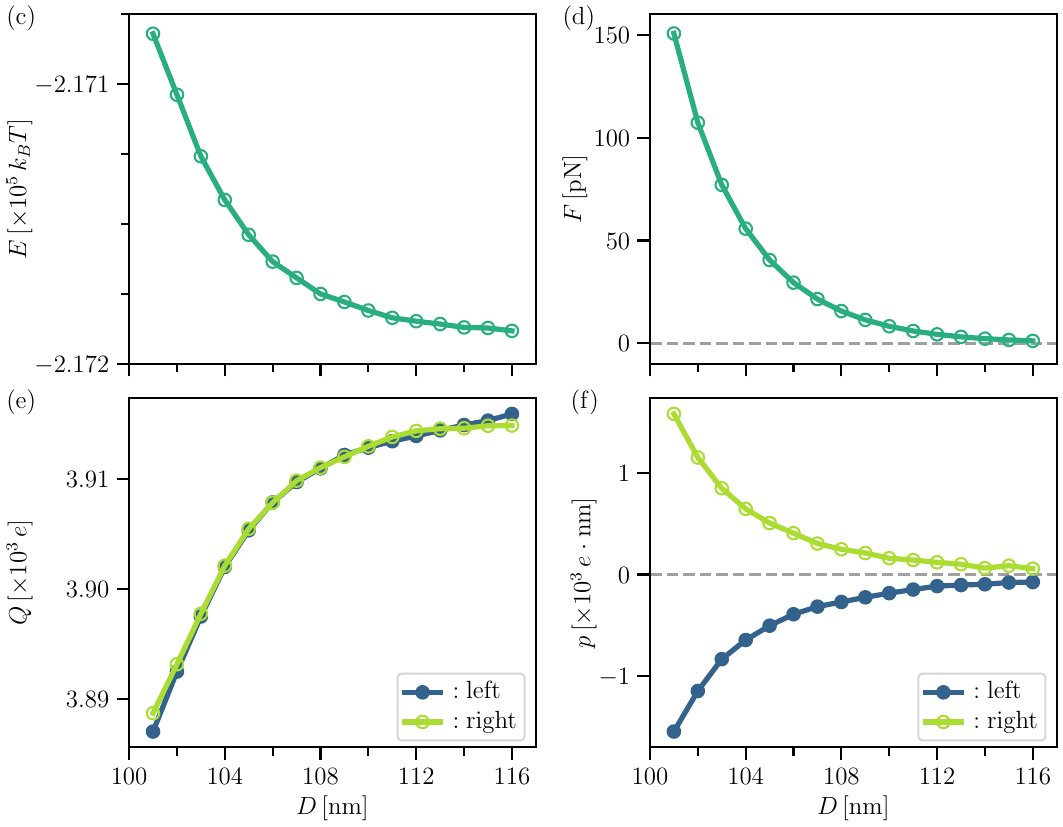}
\caption{Interactions in the Langmuir isotherm dissociation limit with $\chi = 0$ and $\alpha = 3$. Numerical solutions in this case are only symmetric. Surface charge density side view (a), and top view (b), exhibiting a pronounced angular inhomogeneity of the surface charge density stemming from the spherical shape of the macroions.  Total interaction free energy (c) and force (d) as functions of the separation $D$, net charge (e) and dipolar moment (f) for the right (R) and left (L) sphere as functions of the separation $D$. The dipolar moment is a consequence of the nonhomogeneous charge distribution along the macroion surfaces.} 
\label{Fig2}
\end{figure*}
\begin{figure*}[t!]
\includegraphics[width=\textwidth]{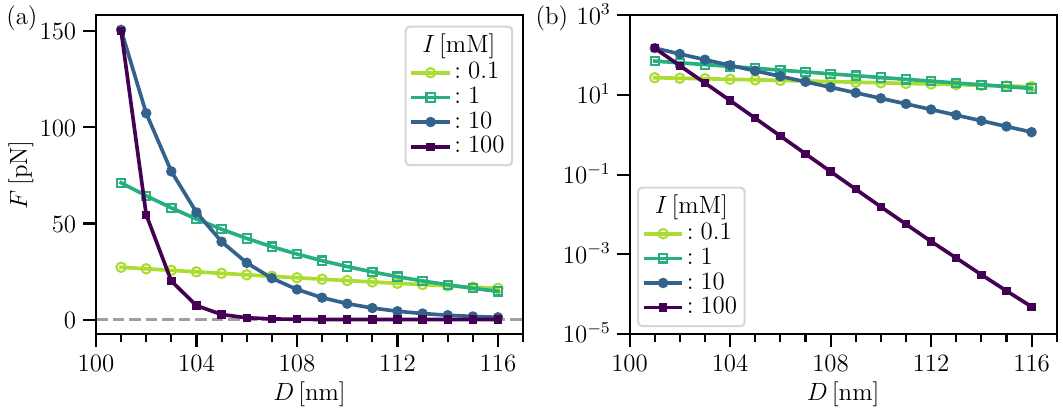}
\caption{The interaction force between two identical macroions of radius 50 nm, as a function of $D$ for $\alpha=3, \chi = 0$ in the linear (a) and log-lin scales (b) for different ionic strengths, $0.1$, $1$, $10$ and $100\,\mathrm{mM}$. The dependence is well characterized by a screened Yukawa potential $e^{-\kappa_D D}/D$ with a standard Debye screening length, Eq.~\eqref{Debyescreen}. The deviations from a linear dependence in the log-lin plot is not seen in the indicated limited range of $D$ values.}
\label{Fig34}
\end{figure*}

\subsection{Interactions and surface charge density}

We first analyse the simplest limiting case of the Langmuir isotherm with $\chi = 0$ and $\alpha = 3$, see Fig.~\ref{Fig2}, for two identical spherical macroions with dielectric cores.

In this case the problem reduces to the Langmuir adsorption of correlated microions with correlations governed by the PB equation. 
The surface charge density is strongly inhomogeneous and has a mirror symmetry with respect to the mid-plane between the macroions (Fig.~\ref{Fig2}a) and shows a pronounced angular dependence (Fig.~\ref{Fig2}b), varying with the separation between them, with the smallest charge density at the two proximal regions. The free energy (Fig.~\ref{Fig2}c) is a decaying function of the separation $D$ and corresponds to repulsive (positive) forces (Fig.~\ref{Fig2}d). 
The total surface charge (Fig.~\ref{Fig2}e) on the two macroions is identical and increases with separation $D$. The angular charge density inhomogeneity engenders also a non-vanishing anti-symmetric dipolar moment whose dependence on the separation is depicted in Fig.~\ref{Fig2}f. 

A detailed look at the separation dependence of the interaction force, see Fig.~\ref{Fig34}, in the limiting case of the Langmuir isotherm clearly shows that the most important functional dependence is well described by an exponential dependence on the separation between macroions, with the standard Debye screening length. The dependence is actually characterized by a screened Yukawa potential $e^{-\kappa_D D}/D$ with a standard Debye screening length, Eq. \eqref{Debyescreen}, but the deviations from a linear dependence in the log-lin plot is not seen in the indicated range of macroion separation  values.

We next analyze the full Frumkin-Fowler-Guggenheim model with $\alpha = -10$ and $\chi = 20$. 
The numerical procedure to solve the PB equation with the CR boundary conditions starts with an initial guess for the distribution of the electrolyte ions and charges on macroion surfaces. We find that, unlike in the Langmuir case, the numerical procedure in COMSOL Multiphysics package is bi-stable: depending on the initial guess, it settles into one and exactly one of two distinct solutions corresponding to either {\sl asymmetric} (Fig.~\ref{Fig3}a) or {\sl symmetric} (Fig.~\ref{Fig3}b) surface charge distribution on macroions -- referring to the mirror symmetry over the mid-plane.  
We stress that the functional form of each of the two solutions of the PB equation plus boundary conditions {\sl does not depend} on the details of the initial configuration, but only on which "basin of attraction" the initial values are chosen from. The thermodynamically stable solution is then picked according to which one of the two corresponds to a lower  free energy,  itself a functional of the electrostatic potential. This is a simple consequence of the fact that numerically we are solving the PB equation, corresponding to the extremal problem, while the correct minimum has to be obtained from assessing the free energy itself.   

We therefore analyze both solutions of the PB equation with CR boundary conditions separately and compare their stability by evaluating the corresponding free energies, see Fig.~\ref{Fig3}. The dependence of the free energy $\mathcal{E}$ on separation $D$ is different for the two types of solutions, see Fig.~\ref{Fig3}c: it increases with $D$ (attractive forces) for asymmetric solution (shown in Fig.~\ref{Fig3}a for $D = 105$nm) and it decreases with $D$ (repulsive forces) for the symmetric solution (shown in  Fig.~\ref{Fig3}b for $D = 105$nm).  At the critical separation, $D_c = 110\,\mathrm{nm}$ (for this particular values of $\alpha$ and $\chi$), both free energies are the same, the two curves in Fig.~\ref{Fig3}c cross. The asymmetric solution implying attractive interaction is stable for separations $D<D_c$, while for large separations the symmetric solution featuring repulsive forces prevails, in accordance with the intuitive expectation that at larger spacing the effect of charge regulation becomes small and the DLVO paradigm prevails. At the critical separation there is a small but finite discontinuous jump in the interaction force, an {\sl attraction-repulsion crossover}, see Fig.~\ref{Fig3}d. 

The symmetric solution corresponds to equal total charges of the macroions, while the asymmetric solution corresponds to opposite total charges of the two macroions, see Fig.~\ref{Fig3}e. The total dipolar moment, see Fig.~\ref{Fig3}f, is the same but oppositely oriented for the symmetric solution, and equal for the asymmetric solution, see Fig.~\ref{Fig3}f. This implies that the asymmetric solution is therefore actually {\sl antisymmetric}. In both cases the inhomogeneity of the macroion surface charge distribution, giving rise to the total dipolar moment,  decreases with separation $D$ and actually fast becomes too small to be detected. 

The existence of an asymmetric solution, at odds with the standard DLVO paradigm \cite{Ver48}, but ubiquitous in CR models \cite{Maj18, Maj19, Maj20}, is a consequence of the minimization not only of the PB free energy with fixed boundary conditions which always yields symmetric solutions, but also the surface CR contributions, characterized by multiple CR minima. This feature of CR models directly leads to states with broken charge symmetry, and thus to asymmetric solutions.

The transition from attraction at smaller separations $D < D_c$ to repulsion for larger separations is a feature already observed in interacting planar surfaces~\cite{Maj18, Maj19} and in the binary spherical cell model~\cite{Maj20} with the important difference that in the binary spherical cell model, there cannot be any inhomogeneity of the surface charge density as observed in the present calculation of two spherical macroions. In what follows, we make two additional assumptions which will simplify our model and elucidate the reason for the discontinuous change in the interaction force: we approximate the two macroions as point charges, and we evaluate the electrostatic free energy  in the DH approximation by linearizing the PB equation  {(see Appendix \ref{Sec:B-App} for details)}. With these additional provisos certain features of the numerical solution, such as the inhomogeneous charge distribution and finite size of the macroions of course cannot be reproduced, but the key charging mechanism can be treated in this semi-analytical theory -- providing a further insight into the attraction--repulsion crossover.

 {As detailed in Appendix \ref{Sec:B-App}} the simple point charge model nicely reproduces the {\sl symmetry breaking transition}, where, depending on $(\alpha,\chi)$, the interaction force is repulsive in the symmetric and attractive in the symmetry broken regime. For $\alpha = -2$ and large enough $\chi$, the interaction force is a monotonic decaying function of $D$, corresponding to repulsive interactions and symmetric charging of the macroion pair. 
For intermediate $\chi$ the interaction force is a non-monotonic function of $D$, with a discontinuous transition from a repulsive to an attractive branch at smaller separations $D$ -- corresponding also to a discontinuous transition between symmetric and asymmetric charging of the macroion pair.  The position of this symmetry breaking transition  depends on $\chi$.   Clearly at a fixed $\alpha$ as $\chi$ decreases the transition between the attractive/repulsive interaction regimes is displaced towards increasingly larger values of the separation $D$. 

\subsection{Details of the interactions and the charging phenomenology}

So far we presented only the salient features of the interaction between the CR spherical macroions. There are, however, other important details that we address next.

First we take a closer look at the surface charge distribution. The inhomogeneous charge distribution along the two macroion surfaces is connected with their spherical shape and the parameters of the CR model. Notably, in the spherical cell model \cite{Maj20} analysed within the same Frumkin-Fowler-Guggenheim model framework, the equilibrium state has no charge inhomogeneity because in the spherical cell model the presence of the other macroion(s) is "simulated" by the boundary condition on the outer boundary of the cell. Since this boundary is also spherical, the equilibrium state can exhibit a homogeneous charge. However, with two spherical macroions in the interaction geometry of Fig.~\ref{Fig0}, the electrostatic potential is never constant along the surface of the macroion, and since {\sl via} the CR isotherm it determines also the surface charge density it also exhibits inhomogeneity along the macroion surface.

\begin{figure*}[t!]
\raisebox{2cm}
{\includegraphics[width=6.5cm]{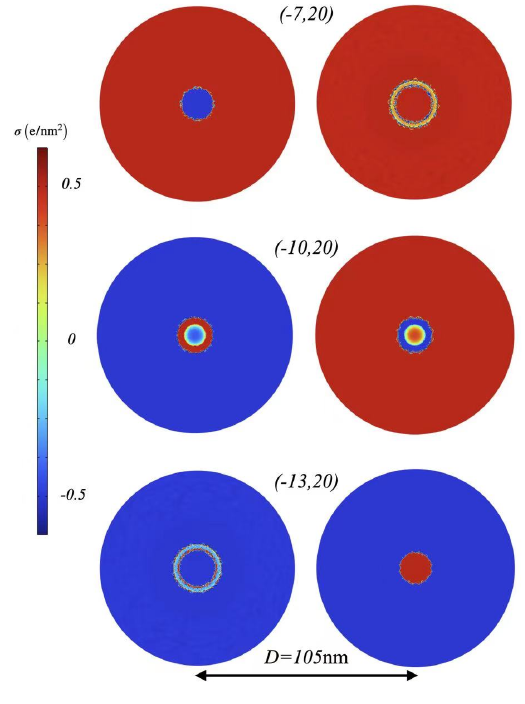}}~~~~~
{\includegraphics[width=10.0cm]{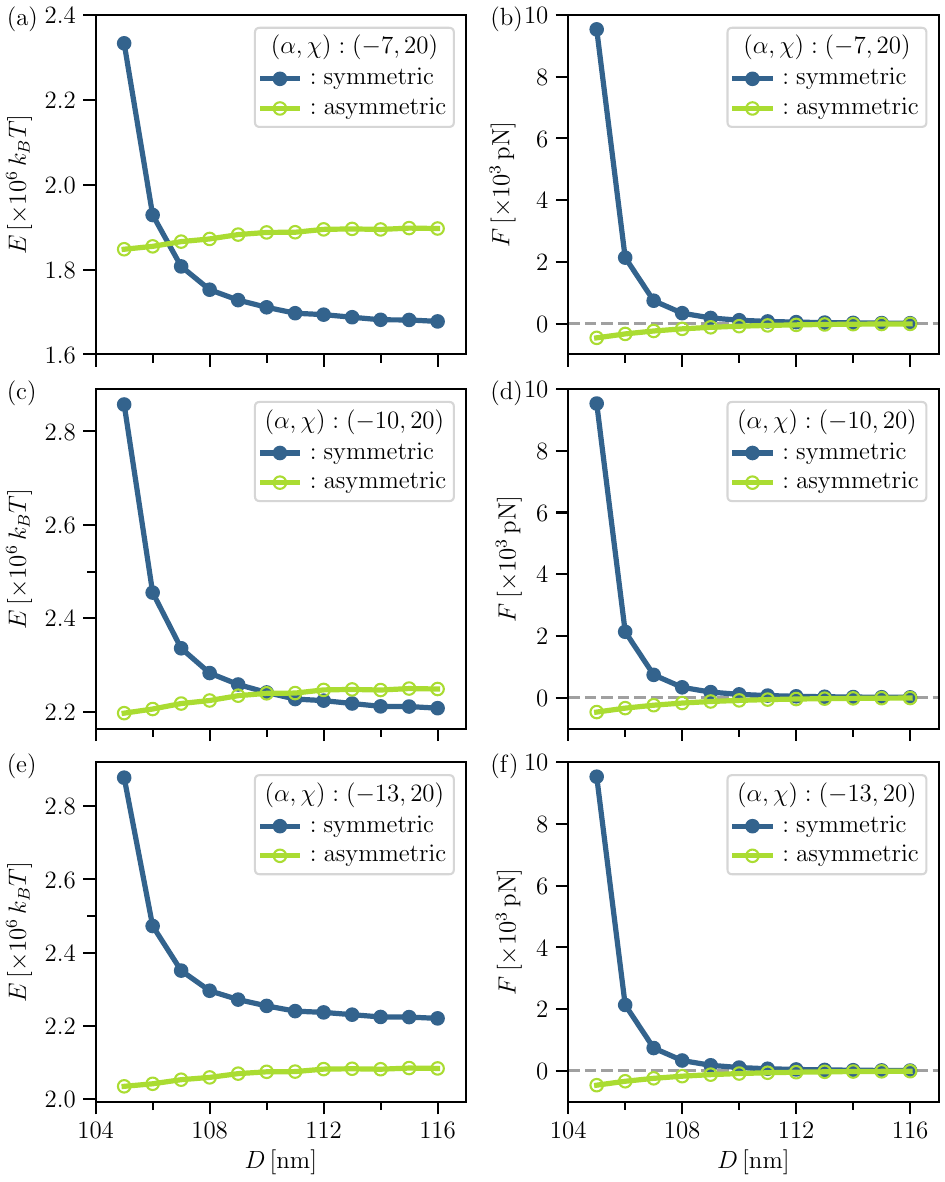}}
\caption{(Left panel) Equilibrium surface charge distribution on the left and right macroion in the axial projection at $D = 105$ nm for three different sets of CR parameters values:   "critical" regime $(\alpha, \chi) = (-10,20)$  "subcritical" regime $(\alpha, \chi) = (-7,20)$ and "supercritical" regime $(\alpha, \chi) = (-13,20)$. In the "critical" regime the solution is antisymmetric, the  charge densities on the two macroions being mirror images. The inhomogeneity is confined to the proximal space between them. In the "supercritical"  regime the charge densities are overall negative, with the asymmetry confined to the proximal region. In the "subcritical"  regime the charge densities are overall positive, with the asymmetry confined to the proximal region. (Right panel) The separation $D$ dependence of the total free energy and the interaction force between the macroions for the same choices of $(\alpha, \chi)$ CR parameters. The separation at which the symmetric and asymmetric solution free energies cross is the critical separation $D_c$ where the corresponding force changes discontinuously from attraction to repulsion. In the "supercritical" regime $D_c$ is displaced way out towards large values of separation.} 
\label{Fig8}
\end{figure*}

In Fig.~\ref{Fig8} we first show the surface charge distribution for three sets of $(\alpha, \chi)$ values pertaining to the "critical" regime, $\alpha = -{\textstyle\frac12}\chi$, the "subcritical" regime of $\alpha \geq -{\textstyle\frac12}\chi$,  and the "supercritical" regime, $\alpha \leq - {\textstyle\frac12} \chi$, always at the same value of spacing $D = 105$ nm. We only describe the equilibrium state, corresponding to the minimal free energy.

For the set with $\chi = 20$, in the "supercritical" regime $(-13, 20)$, we see that while the {\sl total charge} on both macroions is predominantly negative, the {\sl surface charge density} is nevertheless asymmetric, with an inhomogeneity confined to the proximal region of the macroions facing one another.  In the "subcritical" regime $(-7, 20)$  the {\sl total charge} on both macroions is positive, with asymmetric inhomogeneity again confined to the proximal region of the macroions facing one another. Contrary to these two cases in the "critical" regime the two total charges of the macroions are of opposite sign, with the inhomogeneity of the proximal regions being not only asymmetric but actually {\sl antisymmetric}. 

The dependence of the free energy and the interaction force on the separation $D$ provides insights into the force equilibrium. For all cases shown in Fig.~\ref{Fig8} at large enough separation the symmetric charging state, corresponding to the standard PB solution, appears to be universally the stable one, leading to repulsive forces between macroions. This  implies that in this regime the CR dissociation process would not affect the sign of the macroion-macroion interactions, which would then appear as the standard electrostatic repulsion between two identically charged macroions, albeit with an CR dependent net charge. For the set $\chi = 20$ we note that in the "supercritical" regime $(-13, 20)$  the surface charge density inhomogeneity induces a dipolar-like, small in magnitude,  attraction in the whole range of separations. In the "subcritical" regime $(-7, 20)$ except at the very small separations the interaction is predominantly repulsive and the dipolar surface charge inhomogeneity cannot change the sign of the interaction. In terms of the interactions, the "critical" regime $(-10, 20)$ stands in between the "sub" and "supercritical" cases.  
For intermediate values of the  separation, in the "subcritical" regime, there always exists a discontinuous transition into an asymmetric charging state that implies an attractive interaction at smaller separations, turning into a repulsion at larger separations, with a substantial change in the interaction force across the jump. The exact position of this transition depends on the parameter values $(\alpha, \chi)$. Notably, at a fixed $\alpha$ the transition moves towards larger separation values as $\chi$ is increased but eventually saturates for large enough values.

Clearly the CR dissociation process substantially modifies the interaction phenomenology and cannot be fully understood in terms of the standard electrostatic interaction. This is especially true for the emergent attraction in the "supercritical"  regime. 

\begin{figure*}[t!]
\includegraphics[width=16cm]{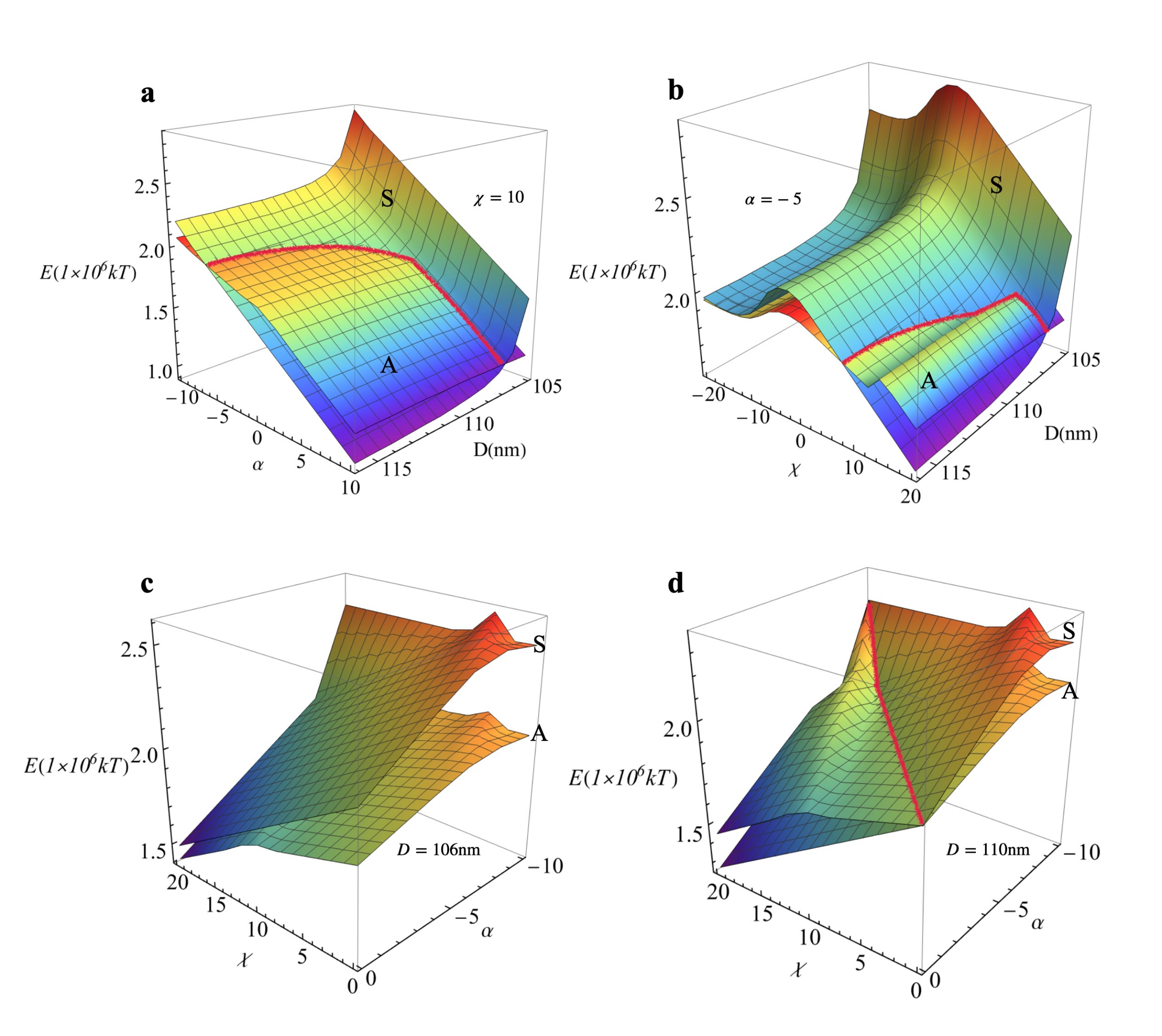}
\caption{Interaction "phase diagram" representation of the nature of interactions between the two colloidal macroions showing free energy as a function of  $D$ and $\alpha$  for a fixed $\chi = 10$ (a) and as a function of $D$ and $\chi$ for a fixed $\alpha = -5$ (b). The line between the "supercritical" and "subcritical" regimes is indicated in red.  In the "supercritical" regime the symmetric solution branch  (marked by "S") is always the stable one, while in the "subcritical" regime at small enough separations there exists a discontinuous transition into an asymmetric solution branch  (marked by "A") that implies an attractive interaction at small separations. The $(\alpha, \chi)$ "phase diagram"  for two different values of the separation, $D=106\,\mathrm{nm}$ (c), and  $D=110\,\mathrm{nm}$ (d). At the smaller separation $D=106\,\mathrm{nm}$ the asymmetric solution (upper surface) is always the stable one (c), whereas at the larger separation $D=110\,\mathrm{nm}$ there is a crossover between the symmetric and asymmetric solution, implying also a change between repulsion and attraction, at the "critical line" $\alpha = - {\textstyle\frac12} \chi$ indicated as the red line (d).} 
\label{Fig7}
\end{figure*}

\subsection{Interaction "phase diagram"}

A more concise representation of the nature of electrostatic interactions in this complicated CR system is provided by the "phase diagram" of Fig.~\ref{Fig7}, showing the asymmetric and symmetric branches of the free energy surface as a function of $(\alpha, \chi, D)$, in the $(D, \alpha, \chi)$, $(D, \chi)$ and $(\alpha, \chi)$ cuts, the latter for two different values of $D$. These free energy surfaces can either cross one another, implying the change in the charge symmetry/nature of the interaction, or remain separate implying that either the symmetric or the asymmetric solution branches stay stable in the indicated range of parameters.

We first analyze the interaction as a function of the separation between the macroions $D$ in the cuts $(D, \alpha)$ and $(D, \chi)$ through the interaction "phase diagram",  Fig.~\ref{Fig7} (a) and (b). For fixed $\chi = 10$, the $(D, \alpha)$ cut shows that the two branches cross along a critical line $D = D_c(\alpha)$. For $D \leq D_c$ the asymmetric (A) branch is stable corresponding to lower free energy and attractive interactions, while for $D > D_c$ the stability is conferred to the symmetric (S) branch corresponding to repulsive interactions. For fixed $\alpha = -5$, the $(D, \chi)$ cut indicates that the stability of the two branches changes at the critical line $D = D_c(\chi)$, so that again for $D \leq D_c$ the asymmetric branch is stable, while for $D > D_c$ the stability is conferred to the symmetric branch.  

The $(\alpha, \chi)$ cut through the "phase diagram" at fixed $D$, Fig.~\ref{Fig7} (c,d), for two different values of the separation, $D = 106, 110$ nm, provides additional insight into the behavior of this CR interacting system. Confining ourselves to the range $-10 \leq \alpha \leq 0$ and $0 \leq \chi \leq 20$, it appears that within this range of parameter values the asymmetric solution is the stable one at smaller separations, $D = 106\,\mathrm{nm}$, while the intersection between the symmetric and the asymmetric solution free energy surfaces at larger separations, $D = 110$ nm, delimits the "subcritical" regime $\alpha \geq -{\textstyle\frac12}\chi$,  and the "supercritical" regime, $\alpha \leq - {\textstyle\frac12} \chi$, corresponding to attractive and repulsive forces. Clearly the intersection is located at the "critical" separatrix $\alpha = - {\textstyle\frac12} \chi$. 

The detailed nature of the interactions as well as the nature of the electrostatic fields between the macroions therefore depends crucially not only on the separation $D$ but also on the position in the $(\alpha, \chi)$ parameter space. For any {\sl finite range} of parameter values there can thus either exist a "critical" separatrix or not, meaning that there can exist a transition between the attractive and repulsive branches of the macroion interaction or it can be absent. While the "critical" separatrix always exists somewhere in the $(\alpha, \chi)$ parameter space only a finite range of these parameters corresponds to a physically meaningful CR model. 

\begin{figure*}[t!]
\includegraphics[width=12cm]{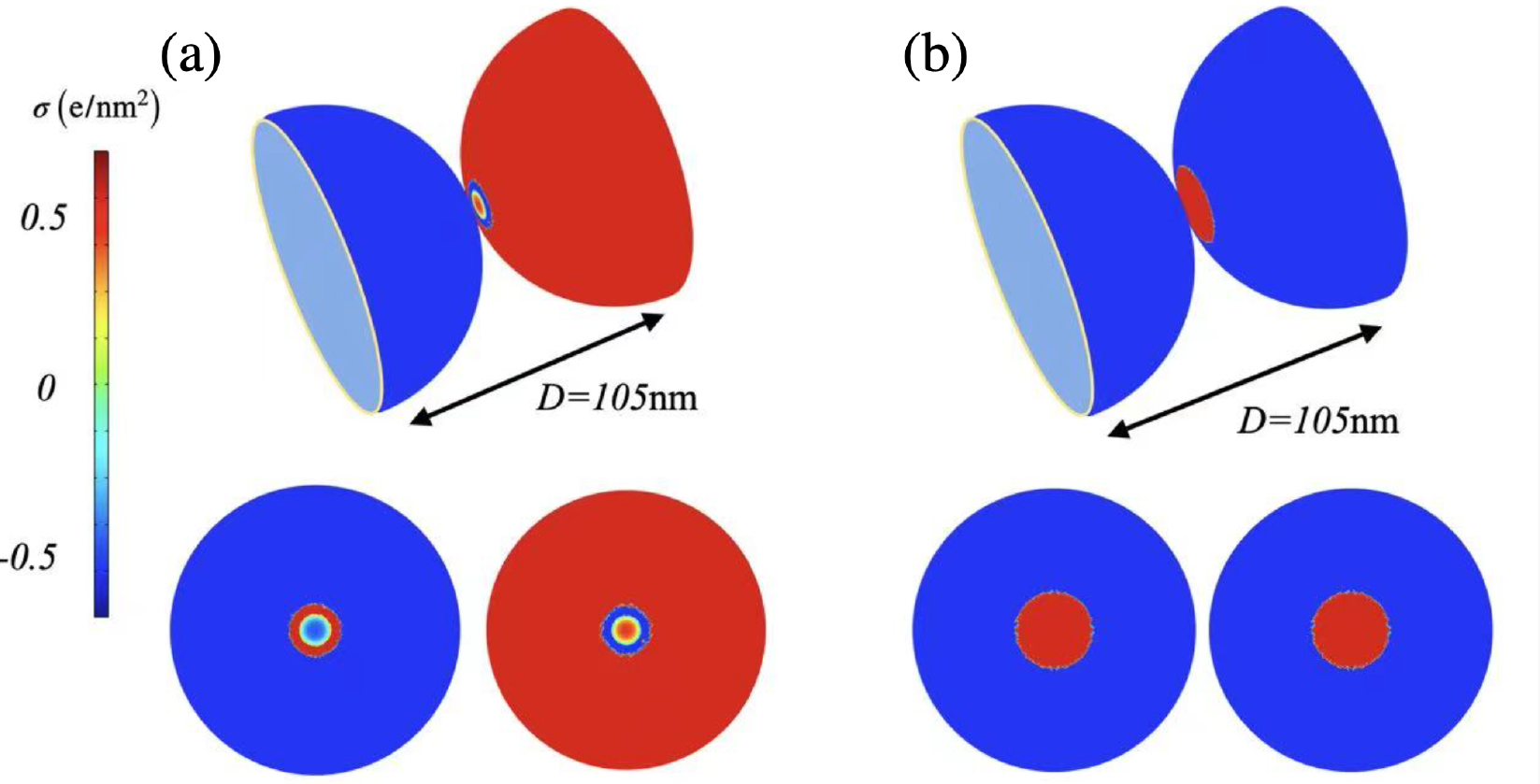}\\
~\\
\includegraphics[width=14cm]{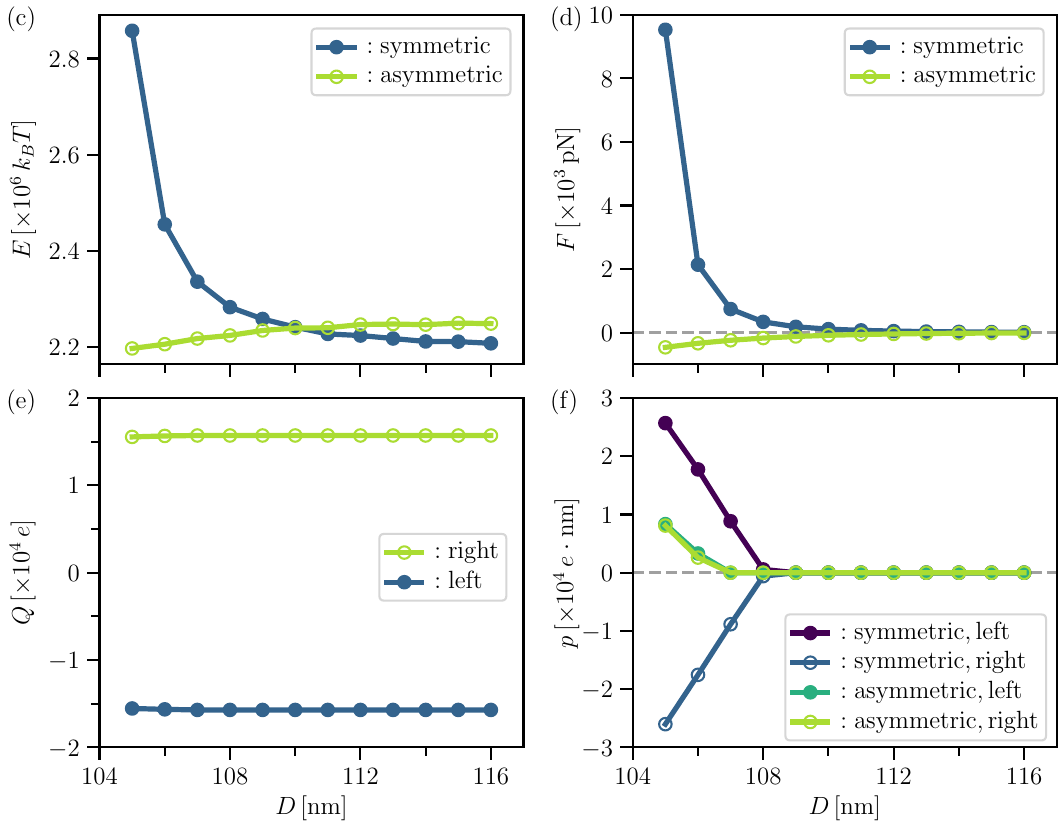}
\caption{Numerical solution for the full Frumkin-Fowler-Guggenheim model with $\alpha = -10$ and $\chi = 20$. Depending on the initial conditions, the numerical solutions in this case are symmetric and asymmetric. (a) surface charge density of the asymmetric solution, side (top) and axis (bottom) view. (b) surface charge density of the symmetric solution, side (top) and axis (bottom) view. Both solutions can be obtained  at separation $D = 105\,\mathrm{nm}$ but the equilibrium solution corresponds to the smaller free energy.  Panels (c), (d), (e) and (f) show free energy, force, total charge, and dipolar moment of the symmetric (S)/asymmetric (A) solutions as functions of the separation. 
Note that the asymmetric solution has a lower free energy for $D \leq 110\,\mathrm{nm}$, whereas the symmetric solution has a smaller free energy for  $D > 110\,\mathrm{nm}$.
The free energies cross at a finite separations ($D = D_c$), denoting a change in the stability of the two solutions, consequently creating a regime of attractive ($D \leq D_c$) and repulsive ($D > D_c$) interactions, with $D_c = 110\,\mathrm{nm}$. The force is therefore non-monotonic and on increase of $D$ changes abruptly from attraction to repulsion as a function of the separation, indicated by dotted line  on panel (d). For the symmetric solution the total charges of both macroions are the same (and consequently cannot be differentiated on the graph), while the dipolar moments are anti-parallel. For the asymmetric solution the two total charges are equal and opposite, while the dipolar moments are identical. The dipolar moments and consequently the inhomogeneity of the surface charge density in general decrease with separation, implying that the inhomogeneity of the surface charge density vanishes for large enough separations.} 
\label{Fig3}
\end{figure*}

\section{Discussion}

Electrostatic interactions between colloidal macroions immersed in a bathing electrolyte solution described within different model frameworks \cite{Muthu2023} have been studied on various levels of approximations, as recently reviewed in detail by Siryk et al. \cite{Siryk2021, Siryk2022}. The weak-coupling \cite{Naji2013} PB equation \cite{Lamm2003}, implying the electric double layer phenomenology \cite{Wu2022,Budkov2022,Huang2023}, is usually taken as a point of departure and the correlation effects extending beyond the mean-field have been addressed either {\sl via} more sophisticated theoretical formulations \cite{Kjellander2020} or directly in simulations \cite{Dijkstra2021}. 

While still remaining within the mean-field level of approximation, a recently growing research direction has focused on a more detailed description of the processes of charge generation itself. In fact, charge dissociation as detailed in the {\sl charge regulation paradigm} \cite{Avn18, Avn19, Avn20} implies that the charges on the macroions are not constant but respond to all the environmental parameters including the separation between the macroions. In this respect we should mention the seminal works of Shklovskii \cite{Shklovski} and Lekner \cite{Lekner} in the limit of {\sl extreme charge regulation}, where the interacting macroions are considered to be conductors with a fixed surface potential. In that case the ubiquitous electrostatic repulsion is almost always turned into attraction at short separations. This leads one straightforwardly to the general question of the possible role of detailed charge regulation in colloid interactions. 

Among the interesting phenomena uncovered recently and also directly related to charge regulation is the {\sl charge symmetry breaking} and its consequences for colloidal interactions \cite{Maj18, Maj19,Maj20}. Here, generalizing the previous charge regulation models \cite{kubincova2020interfacial},  we have applied a Frumkin-Fowler-Guggenheim dissociation isotherm model to evaluate the interactions between two spherical, dielectric macroions with dissociable surface groups, superceding the usual modelling of colloids with fixed surface charge distributions. We have uncovered fundamental modifications in the electrostatic interactions between colloidal macroions and demonstrated that these can strongly depend on the details of the charge generation on macromolecular surfaces immersed in electrolyte solutions.

Our approach is based on a full zwitterionic charge regulation model of macroion surface dissociable groups that allows for positively or negatively charged surface groups, formalized in the framewrok of the Frumkin-Fowler-Guggenheim dissociation isotherm~\cite{Avn18, Avn19, Avn20}. The corresponding CR free energy allows for several local minima which, when coupled to electrolyte mediated electrostatic interactions, leads to possible {\sl symmetry breaking transitions} in the charging of the two spherical macroions~\cite{Maj18, Maj19, Maj20}, implying also a sign change in the forces between the macroions, similar to the case of interactions between CR planar macroions~\cite{Maj18,Maj19, Maj20}. What separates the spherical macroions from planar ones is the emergence of {\sl inhomogeneous charging states}, with surface charge density depending on the position along the interacting surfaces, leading consequently to higher order charge multipoles. While the symmetry broken charge configurations are not universal in this interaction geometry and their emergence hinges on the values of the model CR parameters,  the inhomogeneous charging states are ubiquitous in complex CR systems.

Similar to the case of planar interacting macromolecular surfaces \cite{Maj18,Maj19} we found out that the model CR parameter space is partitioned into two regions, the ”subcritical” regime of $\alpha >{\textstyle\frac12} \chi$ and the "supercritical” regime of $\alpha < {\textstyle\frac12} \chi$, separated by a critical "isotherm"  $\alpha = {\textstyle\frac12} \chi$. In the subcritical regime at $D < D_c$ macroion separations,  the asymmetric charge distribution is the stable one, leading to an attractive interaction, followed by a discontinuous change of stability at $D = D_c$ with the symmetric charge distribution becoming the stable one for all $D > D_c$, where $D_c = D_c(\alpha, \chi)$ is dependent on the parameters of the system. In both the symmetric as well as the asymmetric states the macroion charge distribution is inhomogeneous, described by a net charge and the dipolar moment, to the lowest multipolar order. 
 
We note at this point that the general connection between the surface charging transitions and the nature of interactions between the macroions bears some similarity with the effect that surface ordering transitions have on hydration forces between two surfaces, where the surface free energy exhibits multiple minima, corresponding to different surface ordering states~\cite{Podgornik1989}. Since the Frumkin-Fowler-Guggenheim dissociation isotherm implies a first order dissociation transition, a discontinuous crossover from repulsion to attraction is also consistent with a general theory of the interplay between surface phase transitions and inter-surface forces as formulated for the case of hydration interactions~\cite{Kornyshev1992} as well as simulations within a more simplified model of surface ion adsorption transitions ~\cite{ZHANG1994638}. 

The possibility of asymmetric CR solutions discussed in the present work can be viewed as a novel and alternative source of colloidal attractive interactions in monovalent electrolyte solutions, distinct from either solvation mechanisms ~\cite{kubincova2020interfacial} or strong coupling electrostatics~\cite{Naji2013}, bringing a new paradigm into the study of colloidal interactions between of CR macroions in monovalent electrolyte solutions.   This alternative machanism can be valid in  certain parameter regimes and could shed additional light onto the interplay of observed attractive interactions and charge regulation processes in complex colloids. The proper assessment of the role of this mechanism in colloidal interactions, and its relation to attractive interactions between like-charged macromolecules and surfaces in monovalent electrolyte solutions interpreted as a result of  hydration behavior of molecular water at macromolecular interfaces~\cite{krishnan2007spontaneous,kubincova2020interfacial}, would require a separate probing of the charge density as well as interaction force in order to connect with the CR the theory. Experiments showing anomalous interactions in monovalent electrolyte solutions \cite{Groves2004, Groves2009, Klaassen2022} provide enough motivation to more closely explore the connection between surface charging transitions and interactions.

As for simulations, the CR effects in proteins have a long history \cite{Bar09, Lun13, Barroso_da_Silva2023-ix}, and proceed from a model of the amino acid charge dissociation usually coupled to a model of water self-dissociation. Recently, there has been an additional surge in new methodologies for dealing with charge regulated systems usually based on different types of pH ensembles that are being developed and are available \cite{Radak2017,Lan20, Holm1, Curk2022, Aho2022,Levin2022, Martins_de_Oliveira2022,Buslaev2022}. A specific set of simulations centered around the interaction between charge regulated colloid particles in an electrolyte  \cite{Curk2022, Bakhshandeh2020}, is particularly relevant in the context of our work. Even more specifically, interaction between charge regulated zwitterionic colloid particles \cite{Yuan2022} in a dilute suspension, that showed  a conformational transition from an open assembly of strings or bundles to compact clusters along with the variation in pH,  would be important to investigate in detail to detect possible symmetry changes in the charge distribution. This simulated system could be straightforwardly generalized to include the interaction between two CR macroions as analyzed above. \\
~
\section{Acknowledgements}

HR and RP acknowledge funding from the Key Project No. 12034019 of the National Natural Science Foundation of China and the 1000-Talents Program of the Chinese Foreign Experts Bureau and the School of Physics, University of Chinese Academy of
Sciences. J.D. acknowledges funding from the Chinese National Science Foundation (grants 11874398, 12034019) and from the Strategic Priority
Research Program of the Chinese Academy of Sciences (XDB33000000).

\section{Data availability statement}

The datasets generated during and/or analysed during the current study are available from the corresponding author on reasonable request.

\section{Author Contribution Statement}

R.P and J.D. formulated the problem. R.H. did the calculations. A.M. did the analysis of the results. All the authors contributed equally to the writing of the paper. \\

\appendix

\section{Alternative forms of the free energy \label{Sec:A-App}}

In this appendix we derive alternative forms of the free energy that do not contain the derivatives of the electrostatic potential.

Inserting the Euler-Lagrange equations back into the free energy Eq.~\eqref{equ1} we obtain 
the equilibrium free energy as
\begin{widetext}
\begin{eqnarray}
{\cal E} [\psi({\bf r}), \phi({\bf r})] &=& \frac{k_BT~\kappa_D^2}{4 \pi \ell_B} \!\int_{V}\!\!\!d^3{\bf r}\Big( \frac{1}{2}\beta e_0 \psi  \sinh{\beta e_0 \psi } -  \cosh{\beta e_0 \psi }\Big) +  \sum_{i=1,2}\oint_{S_i}\!\!\!d^2{\bf r}\left( -\frac{1}{2} \psi \frac{\partial f_{CR}(\psi, \phi)}{\partial \psi} + f_{CR}(\psi, \phi)\right).~~~
\label{equfin}
\end{eqnarray}
\end{widetext}
This can be rewritten in a fully symmetric form that as far as we know has not yet been derived in the PB literature. In fact, introducing
\begin{eqnarray}
    f_{PB}(\psi) = - \frac{k_BT~\kappa_D^2}{4 \pi \ell_B} \cosh{\beta e_0 \psi({\bf r})} 
\end{eqnarray}
which is proportional to the (negative) osmotic pressure of the mobile ions, we obtain
\begin{align}
{\cal F} [\psi({\bf r})] &= \int_{V} d^3{\bf r}\Big(-\frac{1}{2} \psi \frac{\partial f_{PB}(\psi)}{\partial \psi} + f_{PB}(\psi) \Big) \nonumber\\ 
&+  \sum_{i=1,2}\oint_{S_i} d^2{\bf r}~ \left( -\frac{1}{2} \psi \frac{\partial f_{CR}(\psi, \phi)}{\partial \psi} + f_{CR}(\psi, \phi)\right).
\end{align}
This form of the free energy is particularly apt for numerical calculations since it contains only the electrostatic potential and the surface charge fraction, but does not contain any derivatives of these fields. We can rework this form of the free energy further by inserting the result of Eqs.~\eqref{surfmin} that yields
\begin{align}
f_{CR}(\psi, \phi) = n_0 k_BT \Big( &- \frac{1}{2}\beta e \psi + \frac{1}{2}\chi \phi^2 \nonumber\\
& - \ln{\left( 1 + e^{- \beta e \psi + \chi \phi + \alpha}\right)}\Big), 
~
\end{align}
a form valid at each of the macroion surfaces $1,2$. From here it also  follows that
\begin{align}
-\frac{1}{2} \psi_i \frac{\partial f_{CR}(\psi, \phi)}{\partial \psi}  = -\frac{1}{2} \psi \sigma = -\frac{1}{2} e n_0~\psi \left( \phi - \frac{1}{2}\right) 
\end{align}
valid at each of the macroion surfaces $1,2$ and therefore combining the two together we remain with
\begin{widetext}
\begin{eqnarray}
&&-\frac{1}{2} \psi \frac{\partial f_{CR}(\psi, \phi)}{\partial \psi} + f_{CR}(\psi, \phi) =  n_0\left( - \frac{1}{2}\beta e \psi +  \frac{1}{2}\chi \phi^2 - \ln{\left( 1 + e^{- \beta e \psi + \chi \phi + \alpha}\right)} -\frac{1}{2}
\beta e\psi \left( \phi - \frac{1}{2}\right) \right). 
\end{eqnarray}
\end{widetext}
again valid at each of the macroion surfaces $1,2$. Inserting this into Eq.~\eqref{equfin} we finally obtain the form of the free energy most suitable for numerical calculations
\begin{widetext}
\begin{align}
{\cal E} [\psi({\bf r}), \phi({\bf r})] 
=& \frac{k_BT~\kappa_D^2}{4 \pi \ell_B} \int_{V} d^3{\bf r}\Big(   \frac{1}{2}~\beta e_0 \psi({\bf r})  ~\sinh{\beta e_0 \psi({\bf r}) } -  \cosh{\beta e_0 \psi({\bf r}) }\Big) \nonumber\\
& + n_0 k_BT \sum_{i=1,2}\oint_{S_i} d^2{\bf r} \Big( - \frac{1}{4}\beta e \psi +  \frac{1}{2}\chi \phi^2 - \ln{\left( 1 + e^{- \beta e \psi + \chi \phi + \alpha}\right)} -\frac{1}{2}
\beta e\psi \phi  \Big).
\label{equfinq}
\end{align}
\end{widetext}
This is a rather simple free energy expression that can be straightforwardly used in numerical calculations  specifically in the context of charge regulation. It remains valid with or without the $\chi$ term, and therefore also in the case of the Langmuir dissociation isotherm for $\chi = 0$. Again, we note that since this free energy contains no derivatives, it is a practical formulation for numerical computation.

\begin{figure*}[hbt!]
{(a)}~~~~~~~~~~~~~~~~~~~~~~~~~~~~~~~~~~~~~~~~~~~~~~~~~~~~~~~~~~~~~~~~~~~~~~~~{(b)}~~~~~~~~~~~~~~~~~~~~~~~~~~~~~~~~~~~~~~~~~~~~~~~~~~~~~~~~~~~~~~~~~~~~~~~~~~~~~~~~~\\
\includegraphics[width=7.5cm]{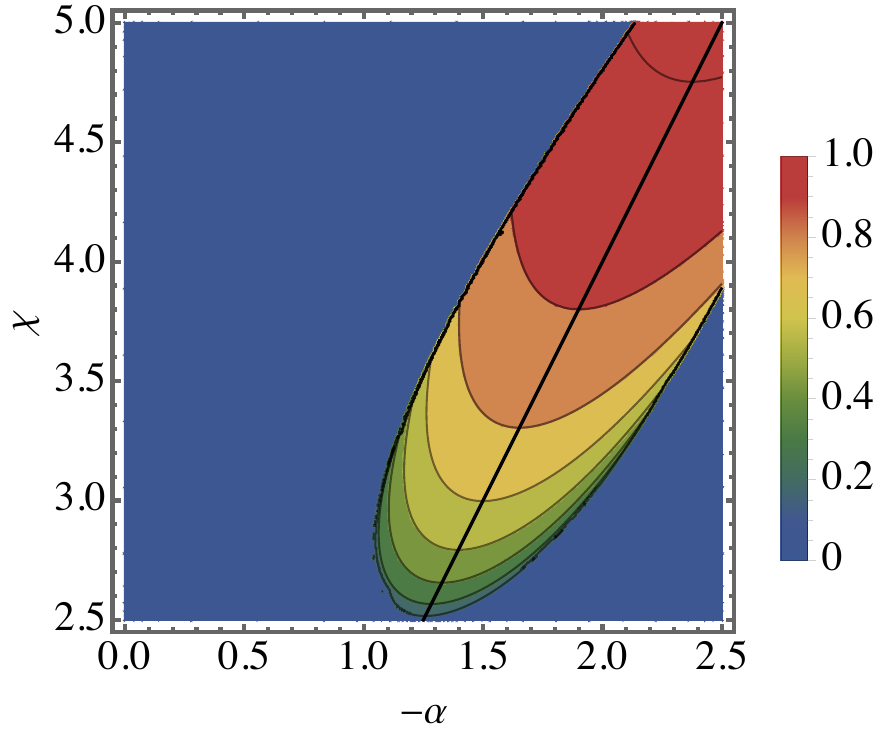}~~~~
\raisebox{4pt}{\includegraphics[width=8.0cm]{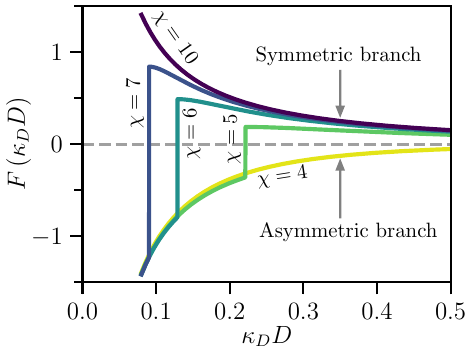}}
\caption{(a) Plot of the charge asymmetry   $\vert\phi_{1}-\phi_2\vert$ for $A = \kappa_D \ell_B/(4\pi)\frac{e^{-\kappa_D D}}{\kappa_D D } = 3$ as a function of $\alpha$ and $\chi$ for two point macroions. The case  $\vert\phi_{1}-\phi_2\vert = 0$, corresponds to a symmetric branch of the solution, while the charge symmetry broken state corresponds to  $\vert\phi_{1}-\phi_2\vert \neq 0$. The line represents the critical dissociation "isotherm" $\alpha = -{\textstyle\frac12}\chi$. 
(b) Plot of the force $-\partial_D{\cal F}(D)$ in the units of $\kappa_D^2 \ell_B/(4\pi)$ as a function of $u = \kappa_D D$. $\alpha = -2$ and $\chi = 10, 7, 6, 5, 4$ (top to bottom curves).  The charge symmetry transitions between the symmetric and asymmetric branches of the solution are now translated into a discontinuous jump in the interaction force  from repulsion to attraction. Note that this discontinuity moves to larger spacing as $\chi$ decreases.} 
\label{Fig33}
\end{figure*}
 
Note that in the absence of CR -- for constant values of the surface charge densities $\sigma_{1,2}$ -- the above free energy reduces to the form equivalent to the one derived by Overbeek \cite{Theodoor1990,Krishnan2017}
\begin{widetext}
\begin{eqnarray}
{\cal E} [\psi({\bf r}), \phi({\bf r})]
 &=& \frac{k_BT~\kappa_D^2}{4 \pi \ell_B} \int_{V} d^3{\bf r}\Big( \frac{1}{2}\beta e_0 \psi({\bf r})  ~\sinh{\beta e_0 \psi({\bf r}) } -  \cosh{\beta e_0 \psi({\bf r}) }\Big)  + \frac{1}{2} \sum_{i=1,2}\oint_{S_i} d^2{\bf r}~ \sigma_i \phi_i.
\label{equfinqfr}
\end{eqnarray}
\end{widetext} 

\section{Point charge limit \label{Sec:B-App}}

We proceed by casting the  electrostatic part of the free energy of the two macroions in the limit of point charges into a much simplified form that can be derived from the Casimir charging process
\begin{align}
\lim_{\left( \oint_{S_1} \!\!\!d^2{\bf r} \longrightarrow 4\pi a^2\right)}\!\!\!\!\!\!\!\!\!\!{\cal F}_{ES}[e_1, e_2] = \int_0^{e_1} \psi(e_1) de_1 +  \int_0^{e_2} \psi(e_2) de_2,
\end{align}
where the limit indicates the point-like macroion approximation, with $e = e_0  ~(\phi-{\textstyle\frac12})$, and the electrostatic potential is given by the DH expression for two point charges separated by $D$, yielding finally the electrostatic free energy in the form
\begin{align}
{\cal F}_{ES} [\psi({\bf r})] \simeq& \frac{e_0^2(\phi_1-{\textstyle\frac12})^2}{4\pi \varepsilon_0\varepsilon_w a} + \frac{e_0^2(\phi_2-{\textstyle\frac12})^2}{4\pi \varepsilon_0\varepsilon_w a} \nonumber\\ 
&+ \frac{e_0^2(\phi_1-{\textstyle\frac12})(\phi_2-{\textstyle\frac12})}{4\pi \varepsilon_0\varepsilon_w} \frac{e^{-\kappa_D D}}{D}. 
\label{DHequ2}
\end{align}
Clearly, the terms linear in $\phi_{1,2}$ simply renormalize $\alpha$ and the terms quadratic in $\phi_{1,2}$ renormalize $\chi$ in the expression for the total free energy, Eq.~\eqref{bgfhjdsk},
\begin{align}
\alpha_{1,2} &\longrightarrow \tilde\alpha_{1,2} = \alpha_{1,2} - \frac{e_0^2}{8\pi \varepsilon_0\varepsilon_w a} - \frac{e_0^2 }{4\pi \varepsilon_0\varepsilon_w} \frac{e^{-\kappa_D D}}{D} \nonumber\\
\chi_{1,2} &\longrightarrow \tilde\chi_{1,2} = \chi_{1,2} +  \frac{e_0^2}{8\pi \varepsilon_0\varepsilon_w a}\, ,
\end{align}
A renormalized and rescaled free energy is then of the form  
\begin{align}
&{\cal F} [\phi_1, \phi_2] \simeq \frac{\kappa_D \ell_B}{4\pi} \frac{e^{-\kappa_D D}}{\kappa_D D }\left(\phi_1-\frac{1}{2}\right)\left(\phi_2-\frac{1}{2}\right)  - \nonumber\\
& - \tilde\alpha_1 \phi_1 - \frac{1}{2} \tilde\chi_1 \phi_1^2 + \phi_1\ln\phi_1+(1-\phi_1)\ln(1-\phi_1)- \nonumber\\
& - \tilde\alpha_2 \phi_2 - \frac{1}{2} \tilde\chi_2 \phi_2^2 + \phi_2\ln\phi_2+(1-\phi_2)\ln(1-\phi_2),
\label{DHequ3}
\end{align}
where $\ell_{B}$ is again the Bjerrum length. One should note the difference between the above free energy and the Langmuir isotherm model used in \cite{Adzic2014,Adzic2015,Adzic2016}. The equilibrium state is obtained numerically -- by minimizing ${\cal F} [\phi_1, \phi_2]$ with respect to $\phi_{1,2}$: 
\begin{eqnarray}
    \frac{\partial {\cal F} [\phi_1, \phi_2]}{\partial \phi_{1,2}(D)} = 0.
\end{eqnarray}
The equilibrium free energy exhibits a separation dependence ${\cal F} [\phi_1(D), \phi_2(D)] \longrightarrow {\cal F}(D)$, and the interaction force is $f=- \partial_D {\cal F}(D)$.

The dependence of free energy on the (dimensionless) separation  $\kappa_D D$ is shown in Fig.~\ref{Fig33}.
Fig.~\ref{Fig33}a displays the charge asymmetry proportional to $\vert\phi_{1}-\phi_2\vert$ as a function of $(\alpha,\chi)$ at fixed $D$. In fact the case  $\vert\phi_{1}-\phi_2\vert = 0$, corresponds to a symmetric branch of the solution, while the   $\vert\phi_{1}-\phi_2\vert \neq 0$ corresponds to charge symmetry broken state. The line in Fig.~\ref{Fig33}a represents the critical dissociation "isotherm" $\alpha = -{\textstyle\frac12}\chi$. 
Clearly, there is an island of asymmetry in the see of symmetric charge partitioning. 
The boundary of this island of asymmetry exhibits either a continuous or discontinuous transition from the symmetric to an asymmetric state, which is reflected in the behavior of ${\cal F}(D)$ in Fig.~\ref{Fig33}b that shows the interaction force dependence on $\kappa_D D$ for fixed $(\alpha,\chi)$. 

\bibliography{manuscript} 
\bibliographystyle{apsrev4-2}

\end{document}